\documentclass[a4paper,fleqn,usenatbib]{mnras}

\usepackage[T1]{fontenc}
\usepackage{ae,aecompl}

\usepackage{graphicx}	
\usepackage{amsmath}	
\usepackage{amssymb}	

\title[Rotation periods and photometric variability of rapidly rotating ultra-cool dwarfs]{Rotation periods and photometric variability of rapidly rotating ultra-cool dwarfs}
\author[P. A. Miles-P\'aez et al.]{
P. A. Miles-P\'aez$^{1,2,3}$\thanks{E-mail: ppaez@uwo.ca},
E. Pall\'e$^{1,2}$,
and M. R. Zapatero Osorio$^{4}$
\\
$^{1}$Instituto de Astrof\'isica de Canarias, Calle V\'ia L\'actea s/n, 38205 La Laguna, Tenerife, Spain\\
$^{2}$Dpt. de Astrof\'isica, Univ. de La Laguna, Avda. Astrof\'isico Francisco S\'anchez s$/$n, 38206 La Laguna, Tenerife, Spain\\
$^{3}$The University of Western Ontario, Department of Physics and Astronomy, 1151 Richmond Avenue, London, ON N6A 3K7, Canada\\
$^{4}$Centro de Astrobiolog\'ia (CSIC-INTA), Carretera de Ajalvir km 4, 28850 Torrej\'on de Ardoz, Madrid, Spain\\
}

\date{}

\pubyear{2017}

\begin{document}
\label{firstpage}
\pagerange{\pageref{firstpage}--\pageref{lastpage}}
\maketitle

\begin{abstract}
 We used the optical and near-infrared imagers located on the Liverpool, the IAC80, and the William Herschel telescopes to monitor 18 M7-L9.5 dwarfs with the objective of measuring their rotation periods. We achieved accuracies typically in the range $\pm$1.5--28 mmag by means of differential photometry, which allowed us to detect photometric variability at the 2$\sigma$ level in the 50\% of the sample. We also detected periodic modulation with periods in the interval 1.5--4.4 h in 9 out of 18 dwarfs that we attribute to rotation. Our variability detections were combined with data from the literature; we found that 65\,$\pm$\,18 $\%$ of M7--L3.5 dwarfs with $v$\,sin\,$i\ge30$ km s$^{-1}$ exhibit photometric variability with typical amplitudes $\le$20 mmag in the $I$-band. For those targets and field ultra-cool dwarfs with measurements of $v$\,sin\,$i$ and rotation period we derived the expected inclination angle of their rotation axis, and found that those with $v$\,sin\,$i\ge30$ km\,s$^{-1}$ are more likely to have inclinations $\gtrsim40$ deg. In addition, we used these rotation periods and others from the literature to study the likely relationship between rotation and linear polarization in dusty ultra-cool dwarfs. We found a correlation between short rotation periods and large values of linear polarization at optical and near-infrared wavelengths.
\end{abstract}




\section{Introduction}\label{sec1}

The presence of magnetic spots and/or dust clouds in the atmosphere of very low-mass stars and brown dwarfs (``ultra-cool dwarfs'', SpT$\,\ge\,$M7) has been proposed to account for the spectro-photometric variability observed by different groups \citep[e.g.,][]{1999MNRAS.304..119T,2001ApJ...557..822M,2002A&A...389..963B,2008MNRAS.391L..88L}. These structures can modulate the shape of the light curve as the dwarf rotates. In some cases, the shape of the observed light curve does not vary significantly with each rotation \citep{2006ApJ...653..690H,2013ApJ...779..101H,2014ApJ...788...23W,2015ApJ...813..104G}, while in other cases, changes in the shape and amplitude are observed after a few rotations \citep{2009ApJ...701.1534A,2012ApJ...750..105R,2013A&A...555L...5G,2014Natur.505..654C}. Infrared observations from space \citep{2014ApJ...782...77B,2015ApJ...799..154M} and the ground \citep{2014ApJ...793...75R} have shown that the amplitudes and occurrence rates of variability do not decrease with spectral type. The observed amplitudes of photometric variability  are typically $\le\,1.5\,\%$ at optical and infrared wavelengths---even though amplitudes $\ge\,2\,\%$ have been observed in some L/T dwarfs---with periodicities in the range $\approx$1.4--20 h, which have been attributed to rotation.  
 
Magnetic activity was one of the first mechanisms proposed to account for the observed photometric variability. H$\alpha$ observations---a sign of magnetic activity---have revealed that nearly 100$\%$ of M7--L3.5 dwarfs are magnetically active \citep{1996AJ....112.2799H,2007AJ....133.2258S,2011AJ....141...97W,2015AJ....149..158S}, so the presence of magnetic spots could lead to photometric variability as in stars \citep{1990ApJS...74..225H,1995AJ....110.2926H,1997A&AS..125...11S}. In addition, optical and radio aurorae have been proposed as another possible origin for some of the variability observed at optical and near-infrared wavelengths \citep{2015Natur.523..568H,2016ApJ...818...24k}. However, strong magnetic fields are not common in ultra-cool dwarfs, especially in those cooler than spectral type L3.5 \citep[only $\sim9\,\%$ of L4--T8 dwarfs are magnetically active;][]{2016ApJ...826...73P}, so magnetic activity alone cannot explain the photometric variability observed from late-M to T dwarfs. 

The low temperatures in the upper atmospheric layers of ultra-cool dwarfs favor the natural formation of condensates \citep[usually referred to as ``clouds'';][]{1996A&A...305L...1T,2001ApJ...556..357A}, which can also induce photometric variability. State-of-the-art radiative transfer models have shown that the observed photometric variability in several ultra-cool dwarfs cannot be explained only by magnetic spots, unless they are combined with dust clouds \citep{2013ApJ...767..173H,2013ApJ...768..121A,2015ApJ...813..104G}. Recently, \citet{2017ApJ...840...83M} investigated a sample of 94 ultra-cool dwarfs, all of them with optical spectra containing the H$\alpha$ region and a photometric monitoring with a duration of at least a few hours. These authors found that the detection rate of magnetic activity and photometric variability follow different dependencies as a function of spectral type, and concluded that $i)$ both phenomena are uncorrelated, and $ii)$ that dust clouds are the most likely origin for the observed photometric variability, even for the warmest ultra-cool dwarfs.
 
The presence of photospheric dust can linearly polarize the output flux of the ultra-cool dwarf by means of scattering processes. This was theoretically studied by \citet{2001ApJ...561L.123S}, \citet{2010ApJ...722L.142S}, \citet{2011MNRAS.417.2874M}, and \citet{2011ApJ...741...59D}. These works predicted values of linear polarization $\lesssim2\%$ at optical and near-infrared wavelengths in good agreement with observations in the filters $R$, $I$, $Z$, $J$, and/or $H$ \citep[see][]{2002A&A...396L..35M,2005ApJ...621..445Z,2009A&A...502..929G,2009A&A...508.1423T,2011ApJ...740....4Z,2013A&A...556A.125M,2015A&A...580L..12M,2017MNRAS.466.3184M}. Non-zero net linear polarization is detectable provided an asymmetry on the surface of the dwarf avoids the cancellation of the polarimetric signal from different areas of the visible disk, such as an oblate shape and$/$or an heterogeneous dust distribution. Both of these asymmetries can be present in ultra-cool dwarfs at the same time; however, their effects on the net polarization are different as the dwarf rotates. For an inhomogeneous atmosphere, the polarimetric signal is expected to modulate with rotation \citep{2011ApJ...741...59D}, while for an asymmetry produced by a non-spherical shape, the most oblate ultra-cool dwarfs show the largest values of linear polarization, but no changes are expected in the signal of linear polarization as the dwarf rotates. \citet{2013A&A...556A.125M} studied the $J$-band linear polarimetric properties of a sample of rapidly rotating ultra-cool dwarfs ($v$\,sin\,$i\ge30$ km\,s$^{-1}$), finding that dwarfs with $v$\,sin\,$i\ge60$ km\,s$^{-1}$ show statistically larger values of linear polarization than those with smaller $v$\,sin\,$i$.

Here, we present the results of our optical ($I$ and SDSS-$i$) and near-infrared ($J$) photometric monitoring of a sample of 18 M7-L9.5  rapidly-rotating ultra-cool dwarfs by using the Liverpool, the IAC80, and the William Herschel telescopes. Our sample is presented in Section \ref{sec2}, and our data acquisition and reduction are described in Section \ref{obs}. In Section \ref{methods}, we describe how the differential intensity light curves of the targets were derived, and how we searched for periodic photometric variability. In Section \ref{sec3}, the rotation periods (T$_{\rm rot}$) found for our sample are compared with others from the literature, and we use them to derive the likely inclination angle of the rotation axis of our targets. Finally in Section \ref{sec4}, we investigate the likely relationship between short rotation periods and large values of linear polarization for some of our targets that have linear polarimetric data acquired with similar filters. Our conclusions are summarized in Section \ref{sec5}.

\begin{table*}
\caption{List of targets.}
\label{Table1}
\tiny
\centering
\renewcommand{\arraystretch}{1.0}
\setlength{\tabcolsep}{2.5pt}
\begin{tabular}{l c c c c c c c c l}
\hline\hline
Name						&SpT		&$i$				&$J$				&$W1$			&$W2$			&$p^{*}_{J}$$^{\rm a}$   	&$v$\,sin\,$i^{\rm b}$	&$R$		&Ref. \\
							&			&(mag)			&(mag)			&(mag)			&(mag)			& $(\%)$			  	&(km\,s$^{-1}$)			&($10^{-3}\,R_\odot$)               	&	  \\
\hline
2MASS J00192626$+$4614078$^{\rm d}$		&M8			&--				&$12.60\pm0.02$	&$11.26\pm0.02$	&$11.00\pm0.02$	&$0.38\pm0.15$		&$68\pm10$                      &		   	& 2\\ 
LP 349$-$25AB$^{\rm d}$			        &M8$+$M9	&$14.387\pm0.004$	&$10.61\pm0.02$	&$9.30\pm0.02$	&$9.05\pm0.02$	&$0.18\pm0.11$		&$55\pm2$(A),$83\pm3$(B)&$144\pm6$(A), $137\pm6$(B)& 5, 8 \\
                                                		&             		&				&                             	&				&				&					&                   			&$171\pm8$(A), $169\pm9$(B) & 9    \\
2MASS J00452143$+$1634446$^{\rm d}$	&L2			&$17.766\pm0.007$	&$13.06\pm0.02$	&$10.77\pm0.02$	&$10.39\pm0.02$	&$0.00\pm0.11$		&$32.8\pm0.2$			&$150\pm20$		  & 1, 11 \\
2MASS J02281101$+$2537380$^{\rm c}$	&L0			&$18.187\pm0.009$	&$13.84\pm0.03$	&$12.12\pm0.02$	&$11.89\pm0.02$	&$0.35\pm0.14$		&$31.2\pm0.8$			&		   		&  1\\ 
2MASS J02411151$-$0326587$^{\rm d}$		&L0			&$20.504\pm0.046$	&$15.80\pm0.07$	&$13.64\pm0.03$	&$13.26\pm0.03$	&$3.04\pm0.30$		&--			                 &$90\pm20$		 & 11  \\
LP 415$-$20AB$^{\rm c}$			        &M7$+$M9.5	&--				&$12.71\pm0.02$	&$11.41\pm0.02$	&$11.19\pm0.02$	&$0.39\pm0.12$		&$40\pm5$(A),$37\pm4$(B)&$100\pm30$(A), $100\pm30$(B)&  5, 9\\
KPNO--Tau 4$^{\rm d}$                                 	&L0	       		&$20.146\pm0.042$	&$15.00\pm0.04$	&$12.80\pm0.02$	&$12.32\pm0.03$	&--					&$10\pm2$ 	                 &360	   	     	&  7, 10\\ 
2MASS J07003664$+$3157266AB$^{\rm c}$  &L3.5$+$L6	&--				&$12.92\pm0.02$	&$10.68\pm0.02$	&$10.38\pm0.02$	&$0.48\pm0.14$		&$30.1\pm2.0$(A)		&		   		&  1, 3\\
PSO J$140.2308+45.6487$$^{\rm c}$    	        &L9.5		&$22.594\pm0.318$	&$15.22\pm0.05$	&$13.06\pm0.02$	&$12.40\pm0.03$	&--					&--					&		   		&  \\ 
2MASS J11593850$+$0057268$^{\rm c}$	&L0			&$18.445\pm0.040$	&$14.08\pm0.03$	&$12.35\pm0.03$	&$12.05\pm0.02$	&$0.53\pm0.15$		&$71\pm2$			&		   		&  3\\
                                                		&             		&$$				&                             	&$ $				&$ $				& $0.40\pm0.13$		&                   			&		   		&    \\
DENIS$-$P J151016.8$-$024107$^{\rm c}$	&M9			&$16.745\pm0.005$	&$12.61\pm0.02$	&$10.94\pm0.02$	&$10.67\pm0.02$	&--                         		&$30\pm4$ 			&		   		& 4  \\
2MASS J15210103$+$5053230$^{\rm c}$	&M7.5		&$15.858\pm0.004$	&$12.01\pm0.02$	&$10.64\pm0.02$	&$10.42\pm0.02$	&$0.60\pm0.13$		&$40\pm4$			&		   		& 2 \\
2MASS J15394189$-$0520428$^{\rm c}$		&L3.5		&--				&$13.92\pm0.03$	&$12.00\pm0.02$	&$11.74\pm0.02$	&--	 				&$40.1\pm0.8$     		&		   		& 1 \\
LP 44$-$162$^{\rm d}$                            		&M7.5		&--				&$11.45\pm0.02$	&$10.13\pm0.02$	&$9.89\pm0.02$	&-- 					&$39.2\pm5.0$ 		&		   		&  6 \\ 
2MASS J18071593$+$5015316$^{\rm c}$	&L1.5		&--				&$12.93\pm0.02$	&$11.24\pm0.02$	&$10.97\pm0.02$	&$0.67\pm0.11$		&$73.6\pm2.2$			&		   		&  1, 3 \\
2MASS J18212815$+$1414010$^{\rm c}$	&L4.5		&--				&$13.43\pm0.02$	&$10.85\pm0.03$	&$10.48\pm0.02$	&--   					&$28.9\pm0.2$ 		 &		   		&  1    \\
2MASS J20360316$+$1051295$^{\rm c}$	&L3			&--				&$13.95\pm0.03$	&$11.90\pm0.02$	&$11.59\pm0.02$	&$0.18\pm0.16$		&$67.1\pm1.5$			&		   		&  1  \\
2MASS J20575409$-$0252302$^{\rm d}$		&L1.5		&$17.624\pm0.007$	&$13.12\pm0.02$	&$11.26\pm0.02$	&$10.98\pm0.02$	&$0.40\pm0.15$		&$60.6\pm2.0$			&		   		& 1, 3  \\
\hline\hline
\end{tabular}
\begin{minipage}{175.5mm}
Notes: $^{\rm a}$ Values of linear polarization taken from \citet{2013A&A...556A.125M,2011ApJ...740....4Z}, and \citet{2017MNRAS.466.3184M}.\\ $^{\rm b}$ Weighted mean rotational velocity for those targets with more than one measurement. Error bars correspond to the average of individual uncertainties quoted in the literature. \\ $^{\rm c}$ Suspected to be older than 500 Myr (see Section \ref{sec2}). \\ $^{\rm d}$ Suspected to be younger than 500 Myr (see Section \ref{sec2}).\\
References. $v$\,sin\,$i$: (1) \citet{2010ApJ...723..684B}; (2) \citet{2010ApJ...710..924R}; (3) \citet{2008ApJ...684.1390R}; (4) \citet{2003ApJ...583..451M}; (5) \citet{2012ApJ...750...79K}; (6) \citet{2012AJ....144...99D}; (7) \citet{2005ApJ...626..498M}. $R$: (8) \citet{2010ApJ...721.1725D}; (9) \citet{2010ApJ...711.1087K}; (10) \citet{2012MNRAS.423.1775M}; (11) \citet{2014A&A...568A...6Z}
\end{minipage}
\end{table*}
   \begin{figure}
   \begin{center}
   \includegraphics[width=0.49\textwidth]{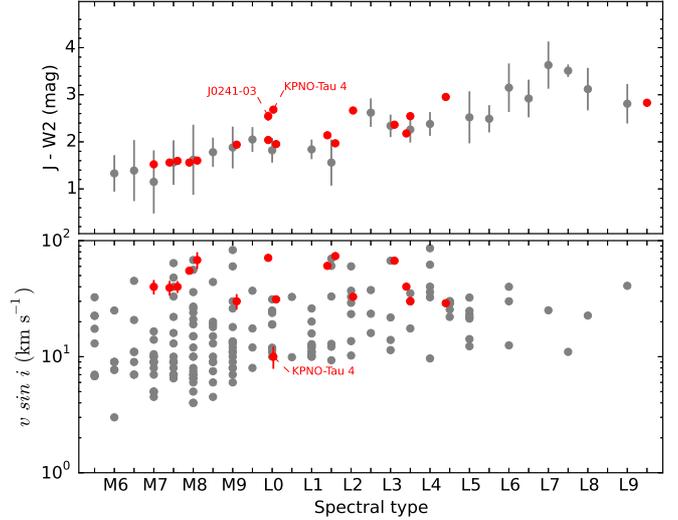}
     \caption{ {\it Top:} {\sl J}--{\sl W2} color of the targets (red circles) as a function of spectral type. We used the average colors reported in \citet{2012ApJS..201...19D} to plot the {\sl J}--{\sl W2} for each spectral subclass with gray circles. Vertical error bars stand for the scatter around the  averaged colors. The error bars associated with our targets are typically smaller than the symbol size. {\it Bottom:} projected rotational velocity ($v$\,sin\,$i$) as a function of spectral type. Our targets are plotted with red circles, while the gray circles are field dwarfs taken from the literature (\citealt{2003ApJ...583..451M}, \citealt{2006ApJ...647.1405Z}, \citealt{2008ApJ...684.1390R}, \citealt{2010ApJ...710..924R}, \citealt{2010ApJ...723..684B}, and \citealt{2012ApJ...750...79K}), their error bars are omitted for clarity. }
              \label{fig0}
     \end{center}
    \end{figure}

\section{Sample description}\label{sec2}

We selected 15 M7--L6 dwarfs with projected rotational velocities ($v$\,sin\,$i$) $\ge\,$30 km\,s$^{-1}$, and 3 L0--L9.5 dwarfs with previously reported photometric variability in the $YJ$ bands \citep[PSO J140.2308$+$45.6487,][]{2013ApJ...777...84B}, or $J$-band linear polarimetric variability \citep[2MASS J02411151$-$0326587 and KPNO--Tau 4, ][]{2017MNRAS.466.3184M}, which suggests the presence of dust clouds. All targets are listed in Table~\ref{Table1},  where we provide their full names (henceforth abridged names will be used), their spectral types, and (if available) their SDSS-$i$, 2MASS-$J$, and WISE magnitudes\footnote[1]{SDSS: Sloan Digital Sky Survey. 2MASS: Two Micron All-Sky Survey. WISE: Wide-Field Infrared Survey Explorer}  \citep{2006AJ....131.1163S,2010AJ....140.1868W,2012ApJS..203...21A,2012yCat.2311....0C,2013yCat.2328....0C}. We also provide published $v$\,sin\,$i$ and $J$-band linear polarization measurements ($p^{*}$) if available, and their references (last column of Table~\ref{Table1}).

In the top panel of Figure \ref{fig0}, we plotted the {\sl J}--{\sl W2} color as a function of spectral type for our sample (red circles) and field dwarfs taken from the literature. As seen in this panel, our targets have colors compatible with those seen in the field, except for the low-gravity objects J0241--0326 (L0) and KPNO-Tau 4 (L0), that are labeled in the figure. In the bottom panel of Figure \ref{fig0}, we plotted the observed $v$\,sin\,$i$ as a function of spectral type. We note that our targets are located in the upper envelope of $v$\,sin\,$i$ for each spectral subclass with the exception of the young brown dwarf KPNO-Tau 4.

\subsection{Age, radius, and binarity}

Based on their spectroscopic properties, in our sample there are 11 targets with likely ages $\ge$500 Myr and 7 objects that are believed to be younger than 500 Myr, including the 1-Myr Taurus member KPNO-Tau 4  \citep{2002ApJ...580..317B}. We identify the young sources in Table \ref{Table1} and summarize the evidence of their youth next: J0019$+$4614 (M8) and J2057$-$0252 (L1.5) show the lithium absorption doublet at 670.8 nm \citep{2008ApJ...684.1390R,2009ApJ...705.1416R} and hallmarks of intermediate-gravity in their near-infrared spectra \citep{2013ApJ...772...79A}; LP 349$-$25AB (M8$+$M9) has an estimated age of $\sim$140--190 Myr \citep[][based on dynamical mass measurements and lack of the lithium absorption]{2010ApJ...721.1725D}; J0045$+$1634 has an age estimation of 10-100 Myr \citep{2014A&A...568A...6Z}  and also was classified as a very-low gravity dwarf by \citet{2013ApJ...772...79A}. J0241$-$0326 (L0) and LP 44-162 (M7.5) have dissenting information on their age. The former was proposed as a member of the Tucana-Horologium moving group \citep[$\sim$20-40 Myr;][]{2014ApJ...783..121G,2015ApJ...798...73G} and the latter was proposed as a member of the ARGUS moving group \citep[$\sim$30-50 Myr;][]{2015ApJ...798...73G}.  However, none of them show lithium in their optical spectra, which points to an age $\ge$500 Myr \citep{2009ApJ...705.1416R,2014A&A...568A...6Z}. 

Published radius estimations are available for 5 targets (LP\,349--25AB, LP\,415--20AB, KPNO-Tau 4, J0045$+$1634, and J0241$-$0326), we list these estimations in column 9 of Table \ref{Table1}. For those targets expected to be older than 500 Myr, theoretical models predict radii in the range $\approx$0.09--0.1 $R_{\sun}$  \citep{1997ApJ...491..856B,2000ApJ...542..464C}. We discuss further the likely size of the targets that exhibit photometric variability in Sections \ref{single} and \ref{binary}.

There are three binaries in our sample (LP 349--25AB, LP 415--20AB, and J0700$+$31AB) with orbital periods of several years \citep{2010ApJ...711.1087K}. By using the average colors of M, L, and T dwarfs listed in \citet{2002AJ....123.3409H} we estimate that the brightness difference between the components is $\Delta\,i\,=\,0.7\pm\,0.3$ mag (LP 349--25AB: M8+M9), $\Delta\,i\,=\,1.3\pm\,0.4$ mag (LP 415--20AB: M7+M9.5), and $\Delta\,i\,=\,1.9\pm\,0.6$ mag (J0700$+$3157AB: L3.5+L6).


\section{Observations and data reduction}\label{obs}

\subsection{Optical data}
We monitored the targets by using the IO:O instrument on the robotic Liverpool Telescope \citep[LT;][]{2004SPIE.5489..679S} in 32 different campaigns, and the imager CAMELOT on the 80-cm IAC80 telescope (IAC80) during 17 campaigns. The total observing time devoted to this project was 112.4 h on the LT, and 76.8 h on the IAC80. The LT and the IAC80 are located on the Observatorio del Roque de los Muchachos (island of La Palma, Spain) and on the Observatorio del Teide (island of Tenerife, Spain), respectively.

IO:O is the optical imaging component of the IO (Infrared-Optical) suite of instruments for the LT. It has a 4096\,$\times$\,4112 pixel E2V CCD 231--84 and a pixel scale of $0.15 \arcsec$, which yields a field of view of $\sim10\arcmin\times10\arcmin$. In all the LT campaigns we used the Sloan-$i$ filter. CAMELOT contains an E2V 2048\,$\times$\,2048 back illuminated chip with $0.304 \arcsec$ pixels$^{-1}$, which corresponds to a $\sim10\arcmin\times10\arcmin$ field of view. We used the Cousin $I$ filter in all the observing campaigns with the IAC80 with the exception of one, in which we collected consecutive measurements in the Sloan-$r$, Sloan-$i$, and Sloan-$z$ filters for J0019$+$4614 during $\sim$5.5 h. The central wavelength and the width of the different filters are: 0.623/0.071 (Sloan-$r$), 0.763/0.076 (Sloan-$i$), 0.966/0.110 (Sloan-$z$), and 0.881/0.156 (Cousin $I$) $\mu$m.

Both instruments (IO:O and CAMELOT) have a linear response up to about 60000 counts. All the targets and the comparison stars, which were used for the differential photometry (see Section \ref{methods}), were far from this limit (typically in the range 1000--30000 counts). Observations were carried out in gray time and always with the targets far from the Moon. Sky was clear in all of the observing campaigns.

Observations in the IAC80 were carried out in visitor mode and targets were typically monitored longer than $\sim$4 h. Observations at the LT were performed in queue mode and to facilitate their scheduling, we required each dwarf to be observed for 8 h split in two observing blocks of 4 h as close in time as possible. Data of both instruments were bias-subtracted, flat-fielded using sky images taken during dusk or dawn, and aligned by using packages within the Image Reduction and Analysis Facility software (IRAF\footnote[2]{IRAF is distributed by the National Optical Astronomy Observatories, which are operated by the Association of Universities for Research in Astronomy, Inc., under cooperative agreement with the National Science Foundation.}).

\subsection{Near-infrared data}
We used the $J$ filter and the Long-slit Intermediate Resolution Infrared Spectrograph (LIRIS; \citealt{2004SPIE.5492.1094M}) on the William Herschel telescope to monitor for $\sim$2 h the L9.5 dwarf PSO J$140+45$, which was reported to vary in the $YJ$-band by \citet{2013ApJ...777...84B}. LIRIS has a 1k$\times$1k HAWAII detector for the 0.8 to 2.5 $\mu$m range, its pixel scale is 0.25$\arcsec$, yielding a field of view of 4.27$\arcmin\,\times\,$4.27$\arcmin$. Data were taken following a nine-point dither pattern with individual exposure times of 60 s, and $x$- and $y$-offsets of 10$\arcsec$. The central wavelength and the width of the $J$ filter is 1.25/0.16 $\mu$m. These data were sky-subtracted by using a sky image obtained from the median combination of groups of 9 images taken at different dithers typically displaced by 10 $\arcsec$, flat-fielded using sky images taken at dusk, and finally aligned using IRAF routines.

In Table \ref{Table2}, we provide the full observing log for the optical and near-infrared data, including: the observing date, the instrument and filter used, number of exposures and individual exposure times, the average seeing of the night, the average signal-to-noise ratio (SNR) of the targets, and the range of airmass during the observations. 

%
%
%
%
\begin{table*}
\caption{Observing log.}
\label{Table2}
\scriptsize
\renewcommand{\arraystretch}{1.0}
\setlength{\tabcolsep}{2.5pt}
\centering
\begin{tabular}{l c c c c c c c c c c}
\hline\hline
Object 			& Obs. date 		&Instrument	&Filter 		&$N\,\times\,t_{\rm exp}$ 	&Seeing	&SNR 	&Comparison stars	&$\sigma_{\rm obs}$     	&$\sigma_{\rm err}$     &Airmass  \\
 				&(UT)			&		        & 			& (s)			                 &(")	        &	      	&				& (mmag)				& (mmag)      		    &	              \\
\hline
J0019$+$4614	&2013 Aug. 11 	&IO:O   		&SDSS$-i$ 	&$66\times200$    			&1.4  	&1390  	&16	 &4.0		 &2.2$\,\pm\,$0.5  		&1.04--1.22 \\
                         	&2013 Aug. 17 	&IO:O   		&SDSS$-i$ 	&$66\times200$    			&1.4  	&1210  	& 16 &5.0		 &2.8$\,\pm\,$0.5  		&1.05--1.30 \\
                         	&2013 Oct. 22 	&CAMELOT   	&Cousins $I$	&$42\times490$ 			&1.7 	&120  	&5 &5.0		 &3.6$\,\pm\,$1.0 		 &1.00--1.52\\
                        	&2013 Nov. 10 	&CAMELOT   	&Cousins $I$	&$45\times360$ 			&1.5 	&331  	&5 &4.0		 &3.9$\,\pm\,$1.5 	&1.05--1.53\\
                        	&2013 Nov. 11 	&CAMELOT   	&Cousins $I$	&$34\times360$ 			&1.4 	&400		& 5 &3.0		 &3.1$\,\pm\,$1.0  		&1.05--1.19\\
                        	&2014 Oct. 02 	&CAMELOT   	&Cousins $I$	&$85\times300$ 			&1.1 	&900  	& 5 &6.0		 &2.5$\,\pm\,$1.0  		&1.05--1.93\\

                        	&2015 Aug. 06 	&CAMELOT   	&SDSS$-r$	&$25\times180$ 			&1.7 	&100  	&13 &38		 &20$\,\pm\,$5 		&1.71--1.09\\
                        	&2015 Aug. 06 	&CAMELOT   	&SDSS$-i$	&$25\times180$ 			&1.6 	&300  	&13 &11		 &4.0$\,\pm\,$1.0  		&1.74--1.08\\
                        	&2015 Aug. 06 	&CAMELOT   	&SDSS$-z$	&$25\times180$ 			&1.6 	&290 	&13 &10		 &4.0$\,\pm\,$1.0  		&1.70--1.08\\

LP\,349$-$25AB	&2013 Oct. 23 	&CAMELOT    	&Cousins $I$	&$68\times300$ 			&1.9 	&300    	&4 &7.0		 &2.0$\,\pm\,$1.0 		&1.00--1.53 \\
J0045$+$1634	&2014 Aug. 10 	&IO:O   		&SDSS$-i$ 	&$45\times300$			&1.6 	&300  	&11 &7.0		 &5.0$\,\pm\,$1.0  		&1.02--1.73 \\
                               &2014 Aug. 24 	&IO:O   		&SDSS$-i$ 	&$45\times300$			&1.2 	&700  &11 &4.0		 &3.0$\,\pm\,$0.5  		&1.02--1.57 \\
                         	&2014 Oct. 14 	&CAMELOT   	&Cousins $I$	&$28\times600$ 			&1.3 	&200  	&10 &5.0		 &4.5$\,\pm\,$1.0  		&1.02--2.11\\
J0228$+$2537	&2013 Oct. 13 	&IO:O   		&SDSS$-i$ 	&$45\times300$    			&1.5 	&100  	&16	 &7.0		 &4.5$\,\pm\,$1.0 		&1.00--1.57\\
                         	&2013 Jan. 24 		&CAMELOT   	&Cousins $I$	&$16\times900$    			&1.4 	&200  	&3 &5.0		 &5.0$\,\pm\,$1.5  		&1.00--1.89\\
J0241$-$0326 	&2014 Sep. 01 	&IO:O   		&SDSS$-i$ 	&$27\times490$    			&1.7 	&80  	&12 &34		 &28$\,\pm\,$3		&1.18--1.87\\
LP\,415$-$20AB 	&2014 Jan. 05  	&IO:O   		&SDSS$-i$ 	&$68\times192$ 			&1.1 	&1290  	&3 &15		 &1.5$\,\pm\,$0.5		&1.01--1.43\\
                        	&2014 Dec. 17 	&IO:O   		&SDSS$-i$ 	&$51\times192$ 			&1.5 	&1000  	& 3 &14		 &2.2$\,\pm\,$1.0  		&1.01--1.17\\
                        	&2015 Jan. 05  	&IO:O   		&SDSS$-i$ 	&$55\times192$ 			&1.2 	&980  	& 3 &14		 &2.3$\,\pm\,$1.0  		&1.01--1.45\\
                          	&2013 Nov. 12 	&CAMELOT   	&Cousins $I$	&$31\times360$ 			&1.2 	&480  	&6 &15		 &4.1$\,\pm\,$0.5  		&1.01--1.11\\
KPNO--Tau 4          &2014 Dec. 18 	&IO:O   		&SDSS$-i$ 	&$26\times490$    			&2.1 	&80  	&6 &44		 &42$\,\pm\,$10 		&1.00--1.41\\
J0700$+$3157AB 	&2014 Jan. 03 		&IO:O   		&SDSS$-i$ 	&$45\times300$ 			&1.2 	&400  	&18 &6.0		 &3.5$\,\pm\,$0.5  		&1.00--1.71\\
                        	&2013 Nov. 10 	&CAMELOT   	&Cousins $I$	&$25\times600$ 			&1.6 	&170  	&24 &5.0		 &5.0$\,\pm\,$0.5  		&1.00--1.17\\
                         	&2013 Nov. 11 	&CAMELOT   	&Cousins $I$	&$32\times600$ 			&1.2 	&350  	& 24 &6.0		 &4.5$\,\pm\,$0.5		&1.00--1.60\\
                        	&2013 Nov. 14 	&CAMELOT   	&Cousins $I$	&$30\times600$ 			&1.2 	&330  	& 24 &5.0		 &4.7$\,\pm\,$0.5  		&1.00--1.58\\
PSO J$140+45$	&2014 Feb. 18 	&LIRIS  		&$J$			&$11\times9\times60$  		&1.1 	&450 	&2 &11		 &8.0$\,\pm\,$3.0     		&1.04-1.06\\
J1159$+$0057	&2014 Feb. 13 	&IO:O  		&SDSS$-i$ 	&$45\times300$  			&1.7 	&180  	&14 &19		 &8.5$\,\pm\,$1.0  		&1.13--1.62\\
                          	&2013 Jan. 25 		&CAMELOT  	&Cousins $I$	&$23\times720$  			&1.4 	&150  	&16 &10		 &10.0$\,\pm\,$1.5  		&1.12--1.85\\
                         	&2015 Apr. 09 	&CAMELOT  	&Cousins $I$	&$40\times420$  			&1.8 	&90  	& 16 &10		 &9.5$\,\pm\,$1.0  		&1.20--1.80\\
J1510$-$0241 	&2014 Mar. 17 	&IO:O   		&SDSS$-i$ 	&$80\times160$ 			&1.3 	&200  	&10 &8.0		 &3.6$\,\pm\,$0.5  		&1.17--1.94\\
                               &2014 May 9    	&IO:O   		&SDSS$-i$ 	&$67\times160$ 			&1.8 	&120  	& 10 &5.0 	 &4.8$\,\pm\,$0.5  		&1.17--1.71\\
J1521$+$5053	&2014 Mar. 16 	&IO:O   		&SDSS$-i$ 	&$80\times160$ 			&1.2 	&700  	&9 &4.0		 &2.3$\,\pm\,$0.5  		&1.10--1.90\\
                               &2014 Mar. 18 	&IO:O   		&SDSS$-i$ 	&$80\times160$ 			&1.1 	&900  	& 9 &3.0		 &2.0$\,\pm\,$0.5  		&1.07--1.30\\
                        	&2013 May 17 	&CAMELOT   	&Cousins $I$	&$32\times360$ 			&1.4 	&500  	&9 &7.0		 &2.6$\,\pm\,$1.0  		&1.12--1.81\\
                        	&2015 May 30 	&CAMELOT   	&Cousins $I$	&$46\times300$ 			&1.4 	&460  	& 9 &4.0		 &4.1$\,\pm\,$1.0  		&1.08--1.84\\
                        	&2015 Jun 04	 	&CAMELOT   	&Cousins $I$	&$49\times300$ 			&1.9 	&200  	& 9 &4.0		 &7.1$\,\pm\,$1.0  		&1.09--1.94\\
J1539$-$0520 	&2014 May 3 		&IO:O   		&SDSS$-i$ 	&$50\times270$  			&1.3 	&200  	&28 &9.0		 &5.0$\,\pm\,$1.0  		&1.20--1.59\\
                               &2014 May 8 		&IO:O   		&SDSS$-i$ 	&$50\times270$  			&1.3 	&160  	& 28 &12.0	 &6.0$\,\pm\,$1.0  		&1.21--1.60\\
LP\,44$-$162	       &2014 Jul. 5 		&IO:O  		&SDSS$-i$ 	&$180\times60$ 	        	&1.3 	&300  	&10 &5.0		 &5.0$\,\pm\,$0.5  		&1.34--1.54\\
                               &2014 Jul. 7 		&IO:O  		&SDSS$-i$ 	&$69\times60$ 	        	&1.0 	&500  	& 10 &4.0		 &3.3$\,\pm\,$0.5  		&1.39--1.51\\
                               &2014 Jul. 10 		&IO:O  		&SDSS$-i$ 	&$180\times60$ 	        	&0.9 	&700  	& 10 &4.0		 &2.8$\,\pm\,$0.5  		&1.37--1.86\\
J1807$+$5015	&2014 May 13 	&IO:O  		&SDSS$-i$ 	&$50\times270$  			&1.3 	&250  	&21 &5.0 	 &3.5$\,\pm\,$1.0  		&1.07--1.26\\
                               &2014 May 17 		&IO:O  		&SDSS$-i$ 	&$50\times270$  			&1.0 	&300  	& 21 &5.0		 &2.5$\,\pm\,$1.0  		&1.07--1.33\\
J1821$+$1414	&2014 May 28 	&IO:O  		&SDSS$-i$ 	&$28\times490$  			&1.5 	&170  	&25 &18		 &8.0$\,\pm\,$2.0  		&1.03--1.25\\
                               &2014 Jun. 1 		&IO:O  		&SDSS$-i$ 	&$28\times490$  			&1.3 	&200  	&25 &7.0		 &5.0$\,\pm\,$1.0  		&1.03--1.20\\
J2036$+$1051	&2013 Jul 17 		&IO:O  		&SDSS$-i$ 	&$28\times490$ 			&1.0 	&250  	&14 &6.0		 &3.7$\,\pm\,$1.0  		&1.05--1.35\\
                               &2013 Jul 30 		&IO:O  		&SDSS$-i$ 	&$28\times490$ 			&1.1 	&220  	& 14 &6.0		 &4.2$\,\pm\,$1.0  		&1.05--1.54\\
                               &2014 Jun. 5 		&IO:O  		&SDSS$-i$ 	&$28\times490$ 			&1.1 	&230  	& 14 &6.0		 &4.1$\,\pm\,$1.0  		&1.10--1.98\\
J2057$-$0252		&2013 Jul. 16  		&IO:O  		&SDSS$-i$ 	&$48\times280$ 			&1.1 	&350  	&12 &3.0		 &3.4$\,\pm\,$1.0  		&1.17--1.50\\
                               &2013 Jul. 31  		&IO:O  		&SDSS$-i$ 	&$48\times280$ 			&1.0 	&330  	& 12 &5.0		 &3.8$\,\pm\,$1.0  		&1.17--1.71\\
                               &2014 Jul. 25  		&IO:O  		&SDSS$-i$ 	&$48\times280$ 			&1.2 	&340  	&12 &4.0		 &3.4$\,\pm\,$1.0  		&1.17--1.71\\
                               &2014 Aug. 25  	&IO:O  		&SDSS$-i$ 	&$48\times280$ 			&1.4 	&340  	&12 &5.0		 &3.5$\,\pm\,$ 1.0 		&1.17--1.68\\
\hline\hline

\end{tabular}

\end{table*}

\section{Analysis}\label{methods}

\subsection{Differential photometry and light curves}

We used VAPHOT \citep{2001phot.work...85D}, which works under the PHOT--IRAF enviroment, to carry out differential photometry. For each aligned image of a time series, its average point spread function was calculated by selecting typically 10-20 stars of brightness similar or greater than the target. Then, we performed circular aperture photometry for all sources in the field of view down to SNR$\,=\,$10, by using apertures of $1\times$ the average Full-Width at Half Maximum (FWHM) computed for each image. We used this aperture to minimize the sky noise contribution in our (typically faint) targets. To obtain the differential intensity light curves of our targets, we measured their fluxes in each image and compared them  with the average flux of a group of comparison stars that were selected in the following manner:

\begin{enumerate}
\item For each instrument, we selected sources with instrumental brightness similar or brighter to the target's, and far from the non-linear regime of the detector. We computed the average value of their measured fluxes, $F_{\rm ref, i}$, in each of the $N$ images of the temporal series ($i=1,\,2...\,N$). The invidual fluxes, $f_{\rm object, i}$, of all of the remaining sources in the field (with the exception of the targets) were compared to this quantity in order to obtain their differential light curves by using the equation:
   \begin{equation}
     \Delta\,m_{\rm object,i} = -2.5\,\log_{10} \left( \frac{f_{\rm object, i}}{F_{\rm ref, i}} \right),\,\,\,\,i=1,\,2,\,...\,N
  \end{equation}
\item We computed the standard deviation ($\sigma$) of all the computed light curves in the previous step, and plotted them against their average differential magnitude ($\Delta\,m$) to pick out those sources with the smallest dispersions. These sources were used to build an optimized set of comparison stars. We obtained the differential light curve of each of these new comparison stars by comparing their fluxes with the average one of the remaining stars of the set. Then, we visually inspected that these light curves were flat, and searched for any periodicity by running a Lomb-Scargle periodogram \citep[LS; ][]{1976Ap&SS..39..447L,1982ApJ...263..835S}. Those stars that showed a significant peak (false alarm probability $<0.1$; FAP) in the periodogram were removed from the budget of selected stable comparison stars valid to perform differential photometry at the time scales of our data. 
\end{enumerate}
 
For each science target we identified between 2 and 28 comparison stars that were used to derive the final differential light curves of all the objects in the field, including the targets. We used the same comparison stars per target in all epochs taken with the same instrument and filter. However for a target observed with different instrumentation, the selected comparison stars data differ from telescope to telescope due to the different exposure times, filters, and telescope sizes. The number of comparison stars used per target and telescope is given in Table\,\ref{Table2}. 

\begin{figure}
   \begin{center}
    \includegraphics[width=0.49\textwidth]{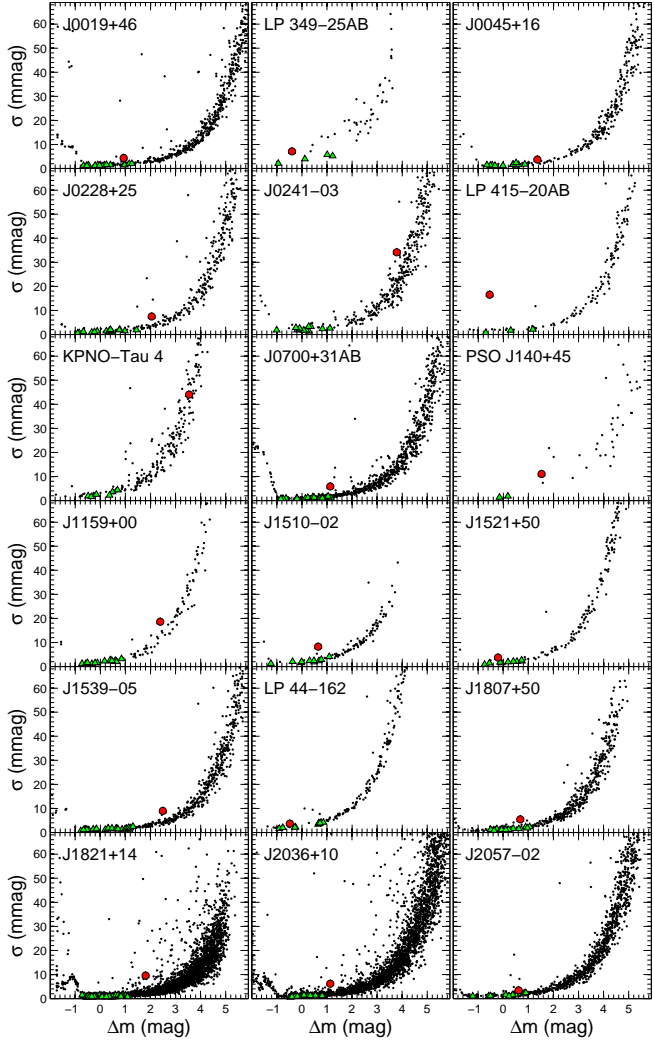}
     \caption{Standard deviation  ($\sigma$) of the differential photometry plotted against the average differential magnitude ($\Delta\,m$) for each object with SNR$\ge10$ in the field of the targets.  We show LT data for all the objects with the exception of LP\,349--25AB (IAC80 data) and PSO J140$+$45 (WHT data). Targets and comparison stars are plotted with red circles and green triangles, respectively.}
              \label{fig0a}
    \end{center}
    \end{figure}

For each source contained in each field, we computed both the average value ($\Delta\,m$) and the standard deviation ($\sigma$) of the time series. These measurements are plotted in Figure  \ref{fig0a}. Sources located in the far left ($\Delta\,m\lesssim-1$ mag) show increasing photometric dispersion as they approximate to the non-linear regime of the detectors, thus these objects are not considered in our analysis due to their poor photometric quality. Sources in the interval $-1\le\Delta\,m\le1$ mag show nearly constant dispersions, yielding the photometric accuracy achieved in our study (typically 1--2 mmag). The selected comparison stars are always located in this photometric interval to guarantee that they do not introduce additional noise in the determination of the target's light curve. The faintest sources ($\Delta\,m>1$ mag) show dispersions that grow as expected for photon noise statistics. The photometric error ($\sigma_{\rm err}$) associated with our targets was determined by taking the average dispersion of objects with similar $\Delta\,m$ ($\approx\pm0.1$ mag), being in the range $\pm$1.5--28 mmag depending on the filter and brightness of the target. In Table\,\ref{Table2}, we provide the dispersion of the light curve of the targets ($\sigma_{\rm obs}$), and their associated photometric error ($\sigma_{\rm err}$) of each epoch. The ratio $\sigma_{\rm obs}/\sigma_{\rm err}$ may be used as a criterion to identify variability: large values of $\sigma_{\rm obs}/\sigma_{\rm err}$ indicate likely photometric variability (see Section \ref{periodicity}).

The differential light curves for each target are plotted in Figures \ref{curvesLT} (LT data) and \ref{curvesIAC} (IAC80 and WHT data). For comparison purposes, in Appendix \ref{ap1}, we include the light curves of one star with similar brightness to the target in Figures \ref{curvesLTref}  and \ref{curvesIACref}. These Figures have the same horizontal and vertical scales as those presented in Figures  \ref{curvesLT} and \ref{curvesIAC}. We also investigated the likely dependence of the differential photometry of our targets with airmass, seeing changes, and their $x$ and $y$ positions on the detectors (before any alignment of the images) to search for any correlation which could lead to false detections or systematic error of variability in the differential intensity light curves. No obvious correlation was found. 

   \begin{figure*}
   \begin{center}
    \includegraphics[width=0.99\textwidth]{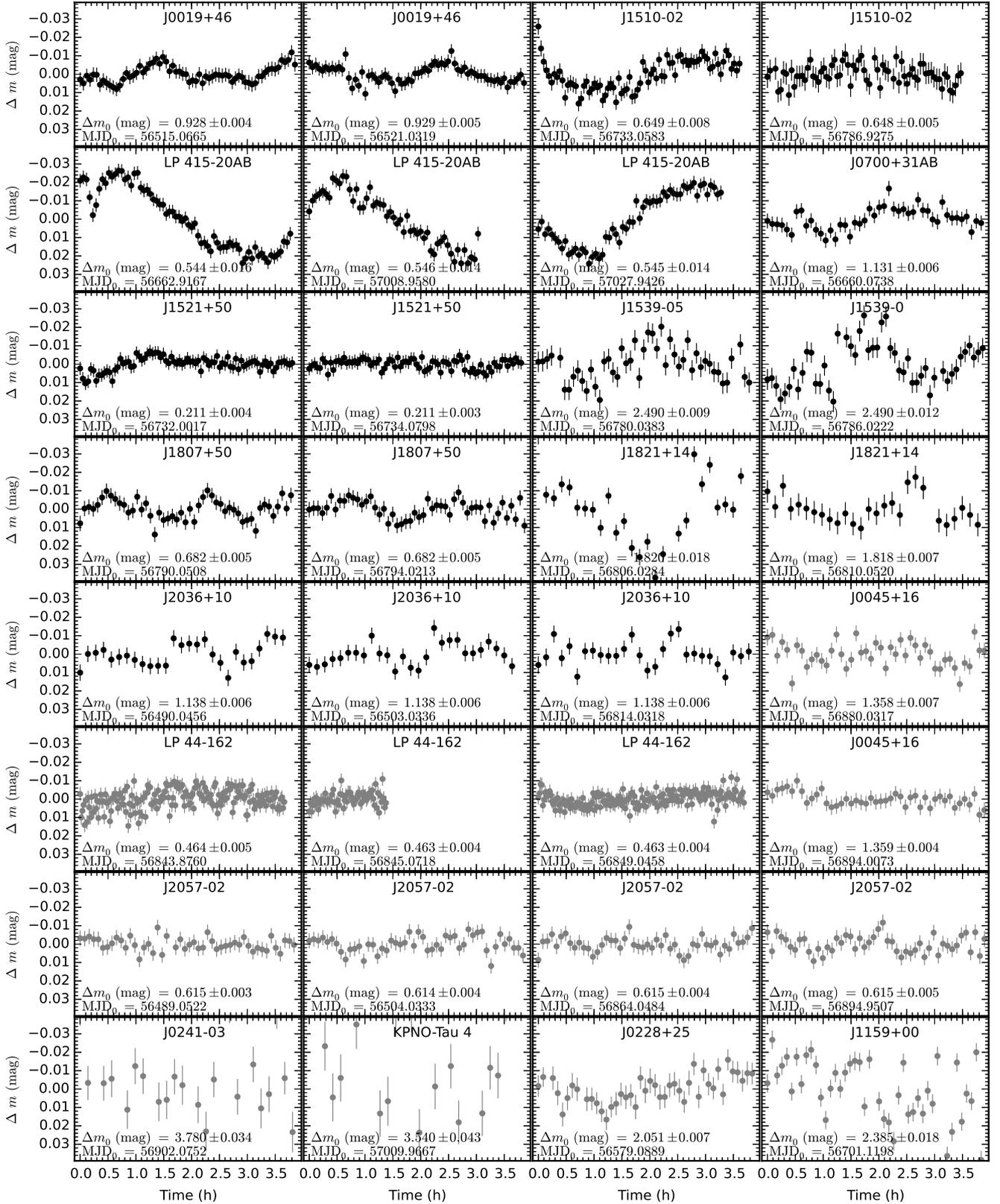}
     \caption{Individual LT light curves for each target (SDSS-$i$ filter), vertical bars stand for photometric errors. The average value of the light curve ($\Delta$m$_{\rm o}$), its standard deviation, and the Modified Julian Date for time zero (MDJ$_{\rm o}$) are also given in all the panels. For comparison purposes, all the light curves have been shifted to the same zero-point, positive and negative values indicate darkening and brightening, respectively. Variable and non-variable objects identified in Section \ref{periodicity} are plotted in black and grey, respectively.}
              \label{curvesLT}
    \end{center}
    \end{figure*}
   \begin{figure*}
   \begin{center}
    \includegraphics[width=0.99\textwidth]{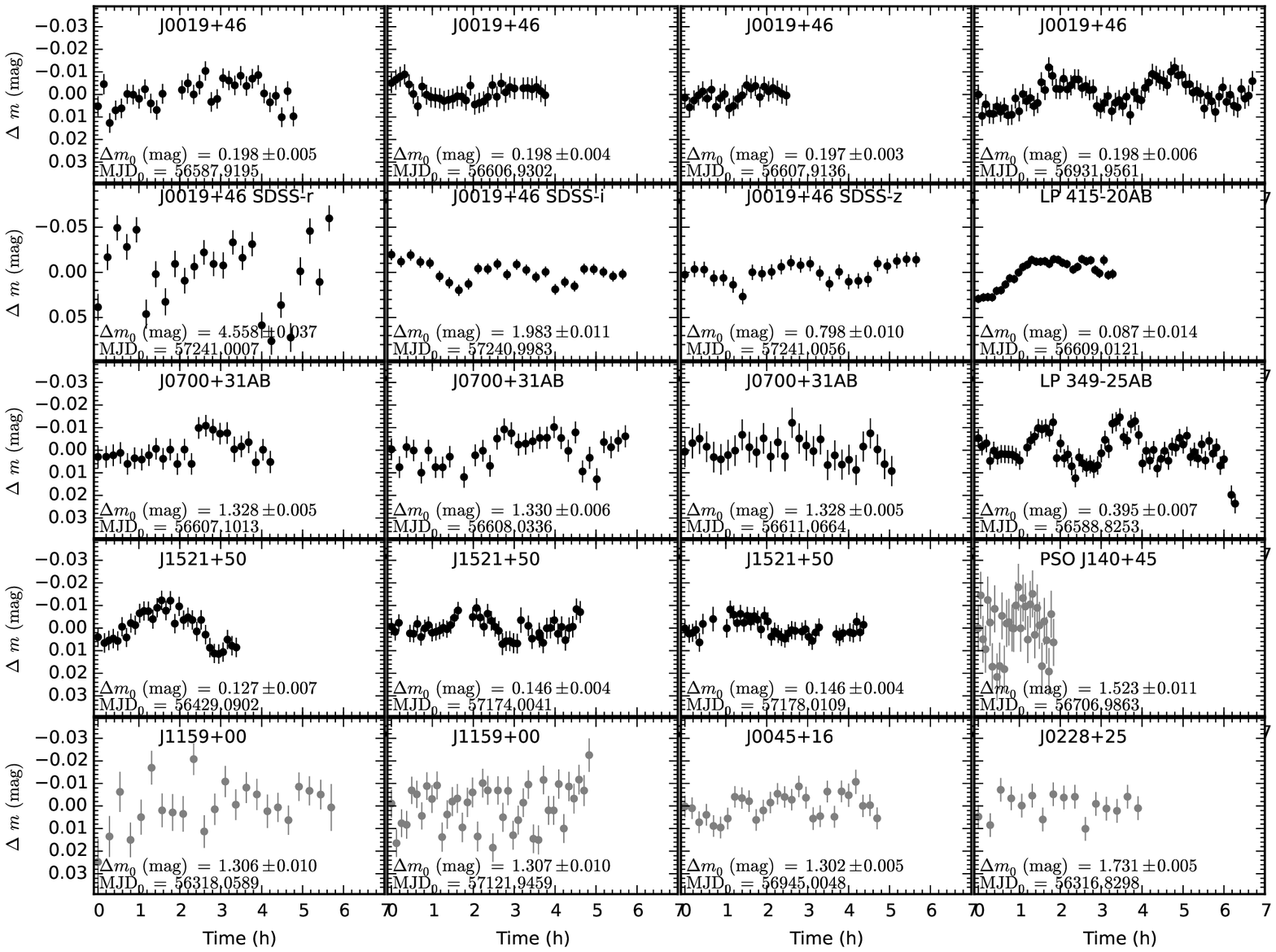}
     \caption{Individual IAC80 light curves for each target. Symbols as explained in Figure\,\ref{curvesLT}. All data were collected in the $I$ filter with the exception of data taken in the Sloan $r$, $i$, and $z$ filters for J0019$+$4614, which are indicated in the corresponding panel, and $J$-band data for PSO J140$+$15. Variable and non-variable objects identified in Section \ref{periodicity} are plotted in black and grey, respectively.}
              \label{curvesIAC}
     \end{center}
    \end{figure*}
%

\subsection{Variability and periodicities}\label{periodicity}

To define photometric variability with the time scales of our study, we use the standard deviation ($\sigma_{\rm obs}$) of the target's light curves shown in Figure \ref{fig0a}. Any object with a dispersion greater than the average value of other sources with similar brightness should be considered as likely variable. We searched for targets whose light curves satisfy $\sigma_{\rm obs}/\sigma_{\rm err}\ge2$ in at least one epoch of observations (columns 9 and 10 in Table \ref{Table2}), and found that 9 out of 18 targets comply with this requirement: J0019$+$4614, LP 349$-$25AB, LP 415$-$20AB, J1159$+$0057, J1510$-$0241, J1521$+$5053, J1539$-$0520, J1807$+$50, and J1821$+$1414. Their light curves can be seen in Figures \ref{curvesLT} and \ref{curvesIAC}; periodic patterns can be easily seen by eye in six of them (J0019$+$4614, LP 349$-$25AB, LP 415$-$20AB, J1521$+$5053, J1539$-$0520, and J1807$+$50), while the remaining 3 (J1159$+$0057, J1510$-$0241, and J1821$+$1414) show stochastic variability.

We searched for variability at low photometric amplitudes in our data by running a LS periodogram since it is able to detect periodic patterns with amplitudes similar to the uncertainty of the data \citep[i.e. $\sigma_{\rm obs}/\sigma_{\rm err}\approx1$; ][]{1986ApJ...302..757H}. We investigated the existence of periodic signals in all the targets independently of whether or not they passed the 2-sigma criterion. 

Besides the LS periodogram, we also used its Bayesian version \citep[BGLS;][]{2015A&A...573A.101M}, and the CLEAN algorithm \citep{1987AJ.....93..968R}. The BGLS provides the relative probability between different peaks of similar power in the LS periodogram, which is useful to discern the most significant peaks in the data. We found that the most significant peaks obtained from the BGLS periodogram are in agreement with those obtained by using the CLEAN algorithm (we used 10-20 iterations and values of 0.1-0.2 for the gain parameter). The LS periodogram is able to detect periodicities in objects with non-sinusoidal light curves, but the retrieved frequency may be far from the real one. Because of this, we also used the phase dispersion minimization technique \citep[PDM; ][]{1965ApJS...11..216L,1978ApJ...224..953S,1990MNRAS.244...93D}. Its basic mechanism relies on the study of which frequency produces the least observational scatter about the mean phase folded light curve of the data.
  
For the LS, BGLS, and PMD algorithms, we explored $10^4$ periods between 2$\times$ the minimum time separation of the data and 16 h ($\sim4\times$ the maximum rotation period expected for objects with $\sim1$ $R_{\rm Jup}$ and $v$\,sin\,$i\ge30$ km\,s$^{-1}$). We visually examined the phase folded light curves of our targets using the most significant peaks retrieved from the algorithms. In all cases, the difference between the most significant peaks obtained by the three algorithms was about $\pm1$ min. 

\begin{figure*}
   \begin{center}
    \includegraphics[width=17cm]{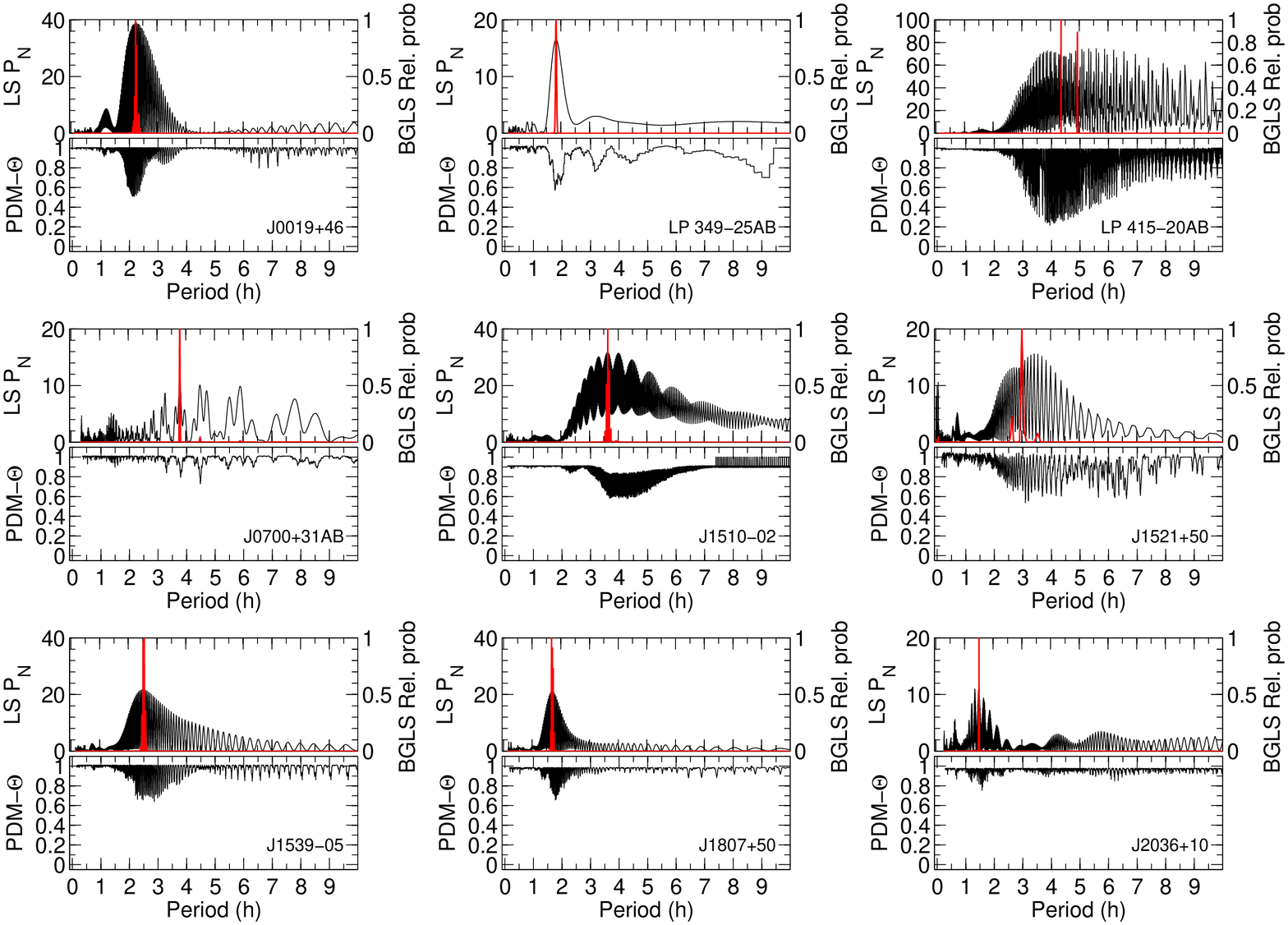}
     \caption{Periodogram algorithms for each of the objects listed in Table\,\ref{Table3}. In each panel, {\sl Top:} Lomb-Scargle (black, vertical axis on the left) and Bayesian Generalized Lomb-Scargle periodogram (red; vertical axis on the right). {\sl Bottom:} Phase Dispersion Minimization algorithm.}
             \label{periodograms}
     \end{center}
    \end{figure*}

We found clear peaks in the periodograms of all targets that passed the 2-$\sigma$ criterion with the exception of J1159$+$0057 and J1821$+$1414. These two dwarfs display stochastic light curves with no obvious patterns nor periodicity. Additionally, the periodograms delivered significant peaks for J0700$+$3157AB and J2036$+$1051, whose light curves have standard deviations  only  12$\%$ above their photometric errors, thus below the 2-$\sigma$ criterion. The periodograms (LS, BGLS, and PMD) for these 9 targets (J0019$+$4614, LP 349$-$25AB, LP 415$-$20AB, J0700$+$3157AB J1510$-$0241, J1521$+$5053, J1539$-$0520, J1807$+$50, and J2036$+$1051) are shown in  Figure\,\ref{periodograms}.

\begin{figure*}
   \begin{center}
    \includegraphics[width=17cm]{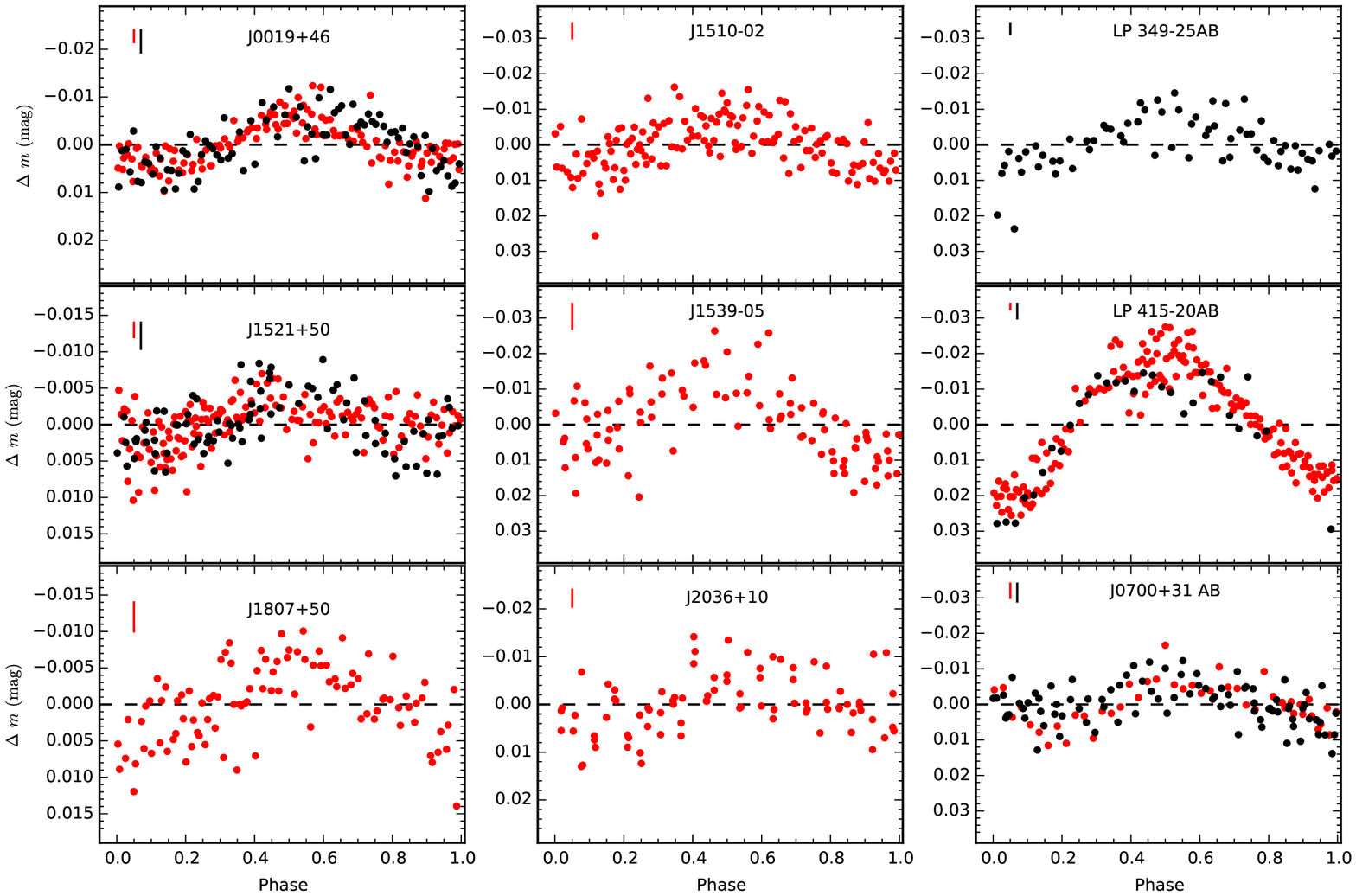}
     \caption{Phase folded light curves for objects of Figure\,\ref{periodograms} using the periods listed in Table\,\ref{Table3}. Data of the Liverpool and IAC80 telescopes are plotted with red and black dots, respectively. Photometric error bars are shown as a vertical bar on the top left side of each panel. Panels on the left and in the middle display the light curves for single objects, and the light curves of the binaries are shown in the panels on the right. Maximum intensity of light curves has been artificially shifted to phase 0.5.}
              \label{phases}
     \end{center}
    \end{figure*}

The significance of the detected peaks was evaluated by using a Fisher randomization test\footnote[2]{We used the False Alarm Probabilities (FAP) defined in the PERIOD-Starlink manual: http://www.starlink.rl.ac.uk/star/docs/sun167.pdf} \citep{1985AJ.....90.2317L}:  assuming that there is no periodic signal in our time series, any of the $\Delta\,m_{i}$ (observed at time $t_{i}$) could have been measured at any other $t_{j}$. Thus, we permuted the original $\Delta\,m_{i}$ sequence to assign them a different time and computed a new LS periodogram on this shuffled time series. We repeated this process $10^{4}$ times for all our targets that showed both a significant peak in their original periodograms and a convincing phase-folded light curve. Then we studied the number of permutations whose peridogram contained a peak larger than the highest peak in the unrandonmized dataset at any frequency. This represents the probability that, given the frequency search parameters, no periodic component is present in the data (FAP1). If a periodic signal is present in the data, the randomization will disturb any relationship between $\Delta\,m$ and time, yielding very small values of FAP1. We also computed the proportion of permutations that contained a peak larger than the peak at the period $P$ in the original data, representing the probability that the period is not actually equal to $P$ but is equal to some other value (FAP2). A peak in the original data was considered as significant if both FAP1 and FAP2 are $\le0.1$. In Table\,\ref{Table3}, we provide the peaks found for the 9 targets and their significance. We used the tabulated peaks of the periodograms to fold in phase the various light curves of each target. The maximum brightness of the light curves was artificially set at phase 0.5. The same period was employed for all light curves corresponding to one object. However, the zero times were conveniently changed for light curves taken on differing epochs (particularly for those observations taken 2 or more days apart: J0019$+$46, J1521$+$50, LP 415--20AB, and J0700$+$31AB) in order to minimize the effects of the error bars associated with the period determination and to maximize the visibility of the periodicity of the folded light curves shown in Figure\,\ref{phases}.
\begin{table*}
\caption{Results of the periodicity search.}
\label{Table3}
\centering
\footnotesize
\renewcommand{\arraystretch}{0.5}
\setlength{\tabcolsep}{1.5pt}
\begin{tabular}{l c c c c c c c r}
\hline\hline
Name			& SpT	&A					&Period				&SIG1$^{\rm a}$	&SIG2$^{\rm a}$	&R\,sin\,$i$     	&$i$  	&$i$ range    \\
				&		&(mmag)				&(h)					&(\%)			&(\%)			&($R_{\sun}$)	&(deg) 	& (deg)     \\
\hline
J0019$+$4614		&M8		&4.7$\,\pm\,$1.5		& $2.22\pm0.41$		&99.9			&99.9 			&$0.124\pm0.017$	& 90		&62-90\\
LP 349$-$25AB 	&M8+M9		&7.0$\,\pm\,$2.0		& $1.83\pm0.20$		&99.9			&99.9 			&--           		&--		&--\\
\,\,\,LP 349$-$25A 	&M8		&-- 					&-- 					&--				&--				&$0.083\pm0.012$ &32		&29-37\\
\,\,\,LP 349$-$25B  	&M9		&-- 					&-- 					&--				&--				&$0.125\pm0.014$ &55		&41-79\\
LP 415$-$20AB       &M7+M9.5		&20.0$\,\pm\,$4.0		&$4.36\pm1.60$ 		&99.9			&99.9			&--      			&-- 		&--\\
\,\,\,LP 415$-$20A    &M7		&-- 					&-- 					&--				&--				&$0.144\pm0.040$ &90		&39-90\\
\,\,\,LP 415$-$20B 	&M9.5	&-- 					&-- 					&--				&--				&$0.133\pm0.040$ &90		&37-90\\
J0700$+$3157A	&L3.5	&4.1$\,\pm\,$2.5		&$3.79\pm1.30$		&99.3			&99.9 			&$0.094\pm0.022$  &81	&52-90\\
J1510$-$0241 		&M9		&6.7$\,\pm\,$2.0		&$3.81^{+1.20}_{-0.70}$  &99.9			&99.9 			&$0.094\pm0.020$ &82		&56-90\\
J1521$+$5053 		&M7.5	&5.0$\,\pm\,$1.5		&$3.09\pm0.9$      		&99.9			&99.9 			&$0.102\pm0.024$  &90	&60-90\\
J1539$-$0520 		&L3.5	&10.5$\,\pm\,$3.0		&$2.51^{+1.60}_{-0.55}$  &99.9			&99.9 			&$0.083\pm0.018$ &61		&43-90\\
J1807$+$5015 		&L1.5	&4.8$\,\pm\,$1.0		&$1.71\pm0.30$ 		&99.9			&99.9 			&$0.104\pm0.012$	 &90		&78-90\\
J2036$+$1051		&L3		&4.2$\,\pm\,$2.0		&$1.45^{+0.55}_{-0.20}$  &99.7			&99.9 			&$0.080\pm0.014$  &58	&45-90\\
\hline\hline
\end{tabular}
\begin{minipage}{175.5mm}
Notes: $^{\rm a}$ The significance, SIG, is defined as $=100\times(1-{\rm FAP})$. FAP1 and FAP2 are described in Section \ref{periodicity}.
\end{minipage}

\end{table*}


\section{Discussion}\label{sec3}

We interpret the peaks of the periodograms to be related to the fast rotation of the targets. According to our determinations, rotation appears to be in the interval 1.2--6.0 h for the ultra-cool dwarfs of our sample with spectral types M7--L3.5 and $v$\,sin\,$i\ge\,30$ km\,s$^{-1}$. The following discussion is based on this interpretation.

The error bars of these periods were estimated as the FWHM measured in the LS periodogram. To determine the amplitude of the periodic light curve, we fitted a sinusoidal curve to the data using as period the peak given by the periodogram. We find that the measured amplitudes of variability (given in Table \ref{Table3}) are comparable to the observed dispersion of the light curves. For the data of the three binaries (LP 349--25AB, LP 415--20AB, and J0700$+$31AB) we subtracted the fitted sinusoidal curve and ran a LS periodogram in the residuals to search for additional periodicities, however, no significant peaks were identified in these periodograms. To the best of our knowledge, this is the first time that a photometric periodicity is reported for: J0019$+$4614, J0700$+$3157AB,  J1510$-$0241, J1521$+$5053, J1539$-$0520, J1807$+$5015, and J2036$+$1051; we confirm the reported variability of LP 349$-$25AB \citep[1.86$\pm$0.02 h, ][]{2013ApJ...779..101H} and  LP 415$-$20AB \citep[4.8 h, ][]{2016ApJ...822...47D}. 

\subsection{Photometric variability}\label{light}
\subsubsection{Single dwarfs}\label{single}
Photometric modulation is found in 6 out of 12 of the targets that do not have any signature of binarity in the literature (J0019$+$4614, J1510--0241, J1521$+$5053, J1539--0520, J1807$+$5015, and J2036$+$1051; Table \ref{Table3}). Additionally, \citet{2015ApJ...799..154M} reported a rotation period of $4.2\pm 0.1$ h for J1821$+$1414, which shows photometric variability at the 2-sigma level in our study (Section \ref{periodicity}) and displays some modulation of $\sim$3--4 h in its light curves of Figure \ref{curvesLT}. However, we failed to find its reported periodicity in our data because it is very close to our typical observing window ($\sim$4 h), and because the shape of its light curve evolves very quickly after a few rotations \citep[][]{2015ApJ...799..154M,2015ApJ...798L..13Y}, which complicates a proper period inference, as seen in our data: the two optical light curves of this object are separated by 4.03 d or $\sim$23 rotation cycles. These changes in the shape of the light curve seem to be common among late L and early T dwarfs \citep[e.g.,][]{2009ApJ...701.1534A,2012ApJ...750..105R,2013A&A...555L...5G}. 

There are several physical phenomena that can produce periodicity. One is the presence of pulsations with predicted periods in the range $\sim$1--5 h due to the deuterium burning at young ages for both low mass stars and brown dwarfs \citep{2005A&A...432L..57P}. However, in our sample only KPNO-Tau-4 is young enough to pulsate due to deuterium burning, and the expected varibility amplitudes due to this mechanism are well in the submmag regime \citep{2013AcA....63...41B,2015MNRAS.446.2613R}, below our data accuracy. Thus, we discard pulsations as the likely variability mechanism of our targets.  

A second source of photometric variability can be the orbit of an unseen companion. However as already mentioned, the targets discussed in this Section are not reported to have massive companions, and their phase-folded light curves in Figure \ref{phases} do not resemble the typical shape of an eclipse.

A third source of variability can be the presence of an inhomogeneous atmosphere that produces photometric modulation as the ultra-cool dwarf rotates. We can evaluate if the observed periods are compatible with the expected rotation velocity of our targets by using the equation:
\begin{equation}
\label{eq1}
    v\,{\rm sin}\,i\,=\,\frac{2\pi\,\times\,R\,{\rm sin}\,i}{\rm T_{\rm rot}}
\end{equation}

where $v$ is the rotational velocity of the surface, R is the radius of the dwarf, and  $\rm T_{\rm rot}$ is the rotation period. With the exception of J0019$+$4614, which has hallmarks of intermediate gravity, the remaining targets are thought to be older than 500 Myr. At these ages  theoretical models predict radii in the range $\approx$0.09-0.1 $R_{\sun}$ \citep{1997ApJ...491..856B,2000ApJ...542..464C}. By using this range of radii, the periods listed in Table \ref{Table3}, and equation \ref{eq1}, we derived rotation velocities in the range 23--37 km s$^{-1}$ (J1510$-$0241), 26--48 km s$^{-1}$ (J1521$+$5053), 26--65 km s$^{-1}$ (J1539$-$0520), 55--80 km s$^{-1}$ (J1807$+$5015), and 54--84 km s$^{-1}$ (J2036$+$1051). All of these velocities are in good agreement with the observed $v$\,sin\,$i$ values of the targets (Table \ref{Table1}). In the case of J0019$+$4614, models predict a radius in the range $\sim$0.11-0.13 $R_{\sun}$ for objects with ages of $\sim$120--500 Myr and effective temperatures typical of M8 dwarfs \citep[$\sim$2500--2600 K; ][]{2009ApJ...702..154S,2013A&A...556A..15R,2013ApJS..208....9P}, this radius and the measured period yield a rotational velocity of 50--80 km s$^{-1}$, which is also in agreement with the reported $v$\,sin\,$i$ value for this object. Thus, we conclude that the observed periodicities of these targets are likely related to rotation. Moreover, the near equality between the estimated rotational velocities and the observed $v$\,sin\,$i$ suggests that the inclination angle of the rotation axis is close to $\sim$90$^{\circ}$, which implies that J0019$+4614$, J1510$-$0241, J1521$+$5053, J1539$-$0520, J1807$+$5015, and J2036$+$1051 are seen face-on. 

\subsubsection{Binaries}\label{binary}

The three binary dwarfs of our sample show photometric variability, which we attribute to rotation. We do not resolve any of the components; therefore, the detected variability resides in the combined flux of both binary members. We do not consider their orbital periods as an explanation for the observed photometric modulation since these are of the order of several years.

We measured a rotation period of  $1.83\pm 0.20$ h for LP 349$-$25AB in agreement with the measurement of $1.86\pm 0.02$ h reported by \citet{2013ApJ...779..101H}. These authors combined their rotation period and the observed $v\,$sin\,$i$ with the assumption of a coplanar spin-orbit alignment to derive the radius of each component of the binary; and compared them to the radii derived from the fitting of the spectral energy distribution of the binary \citep{2010ApJ...721.1725D,2010ApJ...711.1087K}, concluding that LP 349$-$25B was the most likely source of the detected periodicity. Our data alone are not sufficient to evaluate which component is the main responsible for the observed variability. From their brightness difference ($\sim$0.7 mag), we estimate that 65 $\%$ and 35 $\%$ of the observed flux come from LP 349$-$25A and LP 349$-$25B, respectively. If the observed amplitude of variability of the combined flux ($\sim$7 mmag) were produced by LP 349--25B, this would mean that LP 349--25B has an amplitude of variability of $\sim$20 mmag ($\sim$11 mmag for LP 349--25A), which is not very common for objects outside the L/T transition (see Section \ref{comp}). In addition, from the panel of LP 349$-$25AB in Figure\,\ref{fig0a} we would expect typical dispersions of $\sim$2 mmag and $\sim$8 mmag for objects with the brightness of components A and B, respectively. Thus, we can safely measure the variability of LP 349$-$25A, but at the 1-$\sigma$ level we could also measure a modulation coming from LP 349$-$25B. Independently of which companion is originating the variability, the period of $\sim$1.8 h is remarkably stable over 4.08 yr.

In the case of LP 415$-$20AB, \citet{2016ApJ...822...47D} reported a period of $\sim4.8$ h in agreement with our estimation of $4.4\pm1.6$ h (the large error bar is because none of our observations covers one full rotation cycle as seen in Figure\,\ref{curvesLT}). We cannot discern which component is the responsible for the observed variability, or whether it is the combination of both variable components. In Figure\,\ref{phases}, we plotted the phase-folded light curve of LP 415$-$20AB with our period determination. One remarkable feature is that LP 415$-$20AB has the largest variability amplitude ($20\pm4$ mmag) of our sample, which is as large as those observed in some L$/$T transition dwarfs of the field \citep{2012ApJ...750..105R,2013A&A...555L...5G}. From the brightness difference of both components, we estimate that the 75 $\%$ and 25 $\%$  of the total flux come from components A and B, respectively. Thus if only one of the components were variable, we would expect an amplitude of variability of $\sim$26 mmag and $\sim$89 mmag for LP 415--20A and LP 415--20B, respectively. LP 415$-$20AB is thought to be a member of the Hyades cluster \citep[600--850 Myr;][]{2003ApJ...598.1265S,2015ApJ...807...58B}, so both components are expected to have a radius of about 0.1 $R_{\sun}$, as found by \citet[][]{2010ApJ...711.1087K}. These authors also derived an orbital inclination of $55\pm12$ deg for this system. By assuming a spin-orbit alignment and combining the observed $v$\,sin\,$i$ of the components with their radii (see Table \ref{Table1}), we estimate a rotation period of  $\sim$2.5 h and $\sim$2.7 h for components A and B, which are different from the observed 4.4-h periodicity. We could obtain a period closer to the one observed in our light curves by using a radius of $\sim$0.17 R$_\odot$. However, such a large radius is in disagreement with the expected values for the binary components at the age of the Hyades. In the absence of spin-orbit alignment and assuming the radii range expected for the Hyades, we estimate that the maximum rotation period for components A and B is in the range 3.5--3.7 h, which is still in disagreement with the observed modulation. Differential rotation has been suggested to explain photometric periods longer than the expected due to rotation in other ultra-cool dwarfs \citep[BRI 0021$-$0214; ][]{2001ApJ...557..822M},  and it could also explain the periodicity observed in LP 415$-$20AB. 

Finally in the case of J0700$+$31AB (L3.5$+$L6), we estimate the rotation velocity to be 32\,$\pm$\,11 km s$^{-1}$ by assuming a radius of 0.1 $R_{\sun}$ and using the measured rotation period. This rotation velocity estimation is in good agreement with the measured $v$\,sin\,$i$ of  J0700$+$31A. Also taking into account the large $I$-band brightness difference between J0700$+$31A and J0700$+$31B ($\Delta\,I$$\sim$2 mag; $\sim$85 \% of the total flux comes from J0700$+$31A) and the photometric accuracy of our data, we cannot measure variability amplitudes of $\sim$4 mmag for the secondary component, therefore, we attribute the detected rotation period to the component A of the binary. 

\subsubsection{Comparison with the literature}\label{comp}

We have found $I$-band photometric variability in 9 out of 15 (60$\,\pm\,$20 $\%$) ultra-cool dwarfs with $v$\,sin\,$i\ge30$ km s$^{-1}$. One of our targets (LP 349--25AB) was also included in the sample of \citet{2013ApJ...779..101H}. These authors investigated the $R$ and $I$-band photometric variability of 6 radio-detected ultra-cool dwarfs with similar spectral types to our sample, and found periodic modulation in 5 of them; their detection rate and ours are in agreement within 1-$\sigma$, similarly as for our measurements of the amplitudes of variability and theirs. Besides the radio-source LP 349--25AB, other 4 of our variable targets have been observed at 8.46 GHz (J0019$+$4614, J0700$+$3157A, J1521$+$5053, and J1807$+$5015; \citealt{2006ApJ...648..629B,2012ApJ...746...23M}), however, none of them showed radio emission. To the best of our knowledge, the remaining variable targets of our sample (LP 415--20AB, J1510--0241, J1539--0520, and J2036$+$1051) have not been observed at radio wavelengths. All the targets contained in the sample of \citet{2013ApJ...779..101H} are known to have $v$\,sin\,$i\ge30$ km s$^{-1}$. By combining their variability detections and ours, we find that 13 out of 20 (65$\,\pm\,$18 $\%$) M7--L3.5 ultra-cool dwarfs with $v$\,sin\,$i\ge30$ km s$^{-1}$ display optical photometric variability with amplitudes in the range 4--20 mmag. Fast rotators ($v$\,sin\,$i\ge30$ km s$^{-1}$) tend to have large inclination angles (see Section \ref{radius}), so they are more likely to be observed near their equator. This in turn can favor the detection of surface heterogeneities, which produce photometric modulation. 

In the near-infrared, \citet[][]{2014ApJ...797..120R} reported that only 1.4--6\,\% of dwarfs outside the L$/$T transition show variability with amplitudes $\gtrsim20$ mmag, which is in agreement with our findings for warmer dwarfs at optical wavelengths. In addition, the fraction of variable objects of our sample in the optical (60$\,\pm\,$20 $\%$) is also similar to the infrared findings of \citet[][14 out of 23 single L3--L9.5 dwarfs show variability with typical amplitudes of 2--15 mmag]{2015ApJ...799..154M}. Besides the observational bias in our sample (i.e., observations of fast rotators), our detection rate of optical photometric variability for warm ultra-cool dwarfs is similar to the found for cooler dwarfs at infrared wavelengths. 

\begin{table*}
\caption{Objects taken from the literature.}
\label{Table4}
\scriptsize
\centering
\renewcommand{\arraystretch}{1.0}
\setlength{\tabcolsep}{2.5pt}
\begin{tabular}{l c c c c c c l}
\hline\hline
Name						&SpT 	&$p^{*}_{I}$			&$p^{*}_{J}$		&Period				&$v$\,sin\,$i^{\rm a}$	&R\,sin\,$i$		&Ref. \\
							& 		&(\%)				&(\%)			&(h)					&(km\,s$^{-1}$) 		&$R_{\sun}$		&       \\
\hline
BRI\,0021--0214$^{\rm b}$		&M9.5 	&$0.00\pm0.28$		&$0.09\pm0.11$	&$4.8\pm1.0$			&$33.7\pm2.5$		&$0.137\pm0.052$		& 2, 4, 6, 17, 19	\\
LP\,349--25B 					& M8		&$0.27\pm0.37$	&$0.18\pm0.11$	&$1.86\pm0.02$		&$83\pm3$			&--					& 2, 4, 22	\\
2MASS J00361617$+$1821104$^{\rm b}$	&L3.5 	&$0.197\pm0.028$		&$ 0.20\pm0.11$	&$3.08\pm0.05$		&$35.9\pm2.0 $		&$0.091\pm0.007$		& 1, 4, 7, 18	\\
CTI 012657.5$+$280202$^{\rm b}$ 		& M8.5	&$-$				&$-$			&$2.9\pm0.2$			&$11.1\pm2.0$		&$0.027\pm0.007$		& 17, 24	\\
2MASS J01365662$+$0933473$^{\rm b}$	& T2.5	&$-$				&$0.33\pm0.30$	&$2.3895\pm0.0005$	&$-$				&--					& 5, 10	\\
DENIS-P J0255.0$-$4700$^{\rm b}$		& L8		&$0.167\pm0.040$		&$-$			&$1.7\pm0.2$			&$49.4\pm5.0$		&$0.069\pm0.015$		& 1, 12, 17, 18\\
2MASS J03230337$+$4853058	&M6		& $-$				&$-$			&$7.61\pm0.28$		&$33\pm6$			&$0.207\pm0.046$		& 25, 31 \\
2MASS J08283419$-$1309198$^{\rm b}$	& L2		&$-$				&$0.14\pm0.10$	&$2.883\pm0.007$		&$30.1\pm2.0$		&$0.072\pm0.005$		& 4, 8, 18, 20 	\\
							& 		&$-$				&$0.22\pm0.15$	&-- 					&-- 					&					&	\\
2MASS J104842.8$+$011158$^{\rm b}$	&L1 		&$0.00\pm0.10$		&$-$			&$4.71\pm0.10$		&$17\pm2$			&$0.066\pm0.009$		& 2, 14, 20	\\
WISE J104915.57$-$531906.1AB$^{\rm b}$	&L7.5$+$T0.5 &$0.0\pm0.07$	&$-$			&$4.87\pm0.01$		&$26.1\pm0.2$		&$0.105\pm0.005$		& 3, 13, 23 	\\
DENIS-P J1058.7$-$1548$^{\rm b}$ 		&L3 		 &$-$				&$-$			&$4.1\pm0.2$			&$37.5\pm2.5$		&$0.127\pm0.080$		& 16, 26 	\\
Cha H$\alpha$ 2				& M5.25	&$-$				&$-$			&$77.0\pm4.1$		&$12.8\pm2$			&$0.812\pm0.170$		& 27, 28, 29	\\
Cha H$\alpha$ 3				& M5.5	&$-$				&$-$			&$52.6\pm2.2$		&$21\pm2$			&$0.909\pm0.124$		& 27, 28, 29	\\
Cha H$\alpha$ 6				& M5.75	&$-$				&$-$			&$80.6\pm4.6$		&$13\pm2$			&$0.863\pm0.181$		& 27, 28, 29	\\
LHS 2397a$^{\rm b}$ 			&M8 	&$-$				&$-$			&$5.9\pm0.5$			&$23.5\pm2.0$		&$0.114\pm0.019$		& 19, 24	\\
2MASS J11463449$+$2230527$^{\rm b}$ 	&L3$+$L4 &$-$				&$-$			&$5.1\pm0.1$			&$32.5\pm0.2$		&$0.137\pm0.004$		& 17, 31	\\
Kelu-1 AB$^{\rm b}$			& L2$+$L3.5 &$0.75\pm0.27$	&$-$			&$1.80\pm0.05$		&$64.5\pm3.8$		&$0.096\pm0.016$		& 2, 11, 17, 20 \\
2MASS J1334062$+$194034$^{\rm b}$		& L1.5	&$-$				&$-$			&$2.68\pm0.13$		&$25.4\pm4.0$		&$0.056\pm0.012$		& 18, 28	\\
LHS 2924 			 		&M9 	&$-$				&$-$			&$1.8\pm0.2$			&$11\pm2$			&$0.016\pm0.005$		& 17, 24	\\
TVLM\,513--46$^{\rm b}$		& M8.5	&$1.30\pm0.35$				&$0.60\pm0.13$	&$1.959574\pm0.000002$	&$60.0\pm2.0$	&$0.097\pm0.003$		& 4, 9, 17, 32	\\
 							&		&$0.30\pm0.35$				&			&			&			&				&  32	\\
2MASSW J1507476$-$162738$^{\rm b}$ 	&L5		 &$-$				&$-$			&$2.5\pm0.1$			&$21.3\pm2.0$		&$0.044\pm0.017$		& 16, 18	\\
2MASSW J1632291$+$1904407$^{\rm b}$ 	&L8		 &$-$				&$-$			&$3.9\pm0.2$			&$30\pm2$			&$0.096\pm0.037$		& 16, 17	\\
vB 8$^{\rm b}$		 			 	& M7	&$-$				&$-$			&$3.4\pm0.3$			&$9\pm2$			&$0.025\pm0.008$		& 17, 24	\\
DENIS-P J170548.38$-$051645.7$^{\rm b}$	&L4 		&$0.00\pm0.17$		&$-$			&$2.478\pm0.100$		&$27.67\pm0.32$		&$0.057\pm0.003$		& 2, 12, 20	\\
2MASS J17210390$+$3344160$^{\rm b}$	&L3 		&$0.19\pm0.41$		&$-$			&$2.6\pm0.1$			&$-$				&--					& 2,	16 \\
2MASS J18212815$+$1414010	& L4.5	&$-$	&$-$			&$4.2\pm0.1$			&$28.9\pm0.2$		&$0.100\pm0.003$		& 16, 20	\\
2MASS J18353790$+$3259545$^{\rm b}$	& M8.5	&$0.04\pm0.03$	&$0.07\pm0.10$	&$2.84\pm0.01$		&$41.2\pm4.7$		&$0.097\pm0.011$		& 2, 4, 7, 19, 21 \\
2MASS J21041491$-$1037369$^{\rm b}$	&L2.5 	&$0.36\pm0.35$		&$-$			&$1.62\pm0.10$		&$23.44\pm0.23$		&$0.031\pm0.002$		& 2, 8, 20	\\
2MASS J21580457$-$1550098$^{\rm b}$	&L4 		&$1.33\pm0.35$		&$-$			&$1.52\pm0.10$		&$-$				&--					& 2, 8	\\
\hline\hline
\end{tabular}
\begin{minipage}{175.5mm}
Notes: $^{\rm a}$ Weighted mean rotational velocity for those targets with more than one measurement. Error bars correspond to the average of individual uncertainties quoted in the literature. $^{\rm b}$ Likely to be older than 0.5 Gyr.\\
References: {\sl Linear polarization}: (1) \cite{2002A&A...396L..35M}; (2) \citet{2005ApJ...621..445Z}; (3) \cite{2013ApJ...770..124K}; (4) \citet{2013A&A...556A.125M}; (5) \cite{2011ApJ...740....4Z}; (32) \cite{2015A&A...580L..12M}. {\sl Rotation periods}: (6) \cite{2001ApJ...557..822M}; (7) \cite{2008ApJ...684..644H}; (8) \cite{2004MNRAS.354..378K}; (9) \cite{2014ApJ...788...23W}; (10) \cite{2009ApJ...701.1534A}; (11) \cite{2002MNRAS.332..361C}; (12) \cite{2005MNRAS.360.1132K}; (13) \cite{2013A&A...555L...5G}; (14) \cite{2003MNRAS.346..473K}; (15) \cite{2002ApJ...577..433G}; (16) \cite{2015ApJ...799..154M}; (24) \citet{1996ASPC..109..615M}; (25) \citet{1997MNRAS.286L..17M}; (28) \citet{2003ApJ...594..971J}; (30) \citet{2001A&A...367..218B}; (31) \citet{2002A&A...389..963B} . {\sl $v$\,sin\,$i$}: (17) \cite{2003ApJ...583..451M}; (18) \cite{2008ApJ...684.1390R}; (19) \cite{2010ApJ...710..924R}; (20) \cite{2010ApJ...723..684B}; (21) \cite{2012AJ....144...99D}; (22) \cite{2012ApJ...750...79K}; (23) \cite{2014Natur.505..654C}; (26) \cite{2000ApJ...538..363B}; (27) \cite{2001A&A...379L...9J}; (29) \cite{2005ApJ...626..498M}; (31) \cite{1999ApJ...510..266B} 
\end{minipage}
\end{table*}
   \begin{figure}
    \begin{center}
    \includegraphics[width=0.49\textwidth]{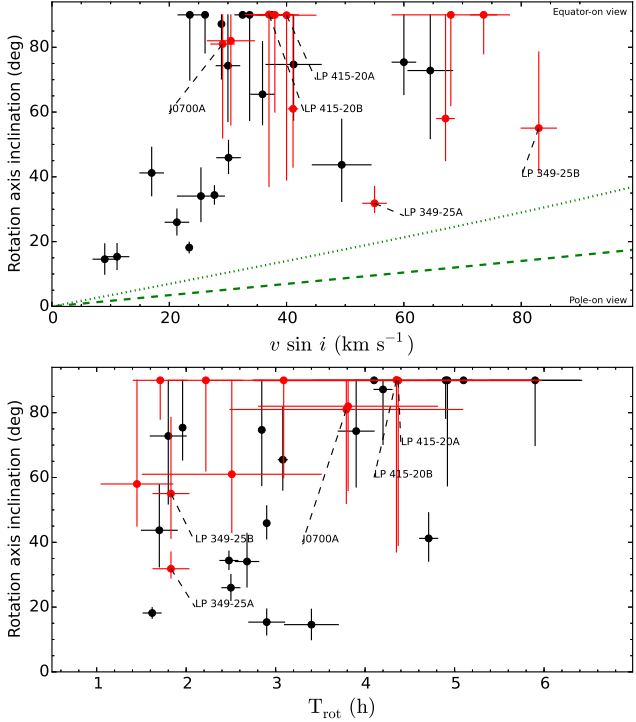}
     \caption{Inclination range as a function of $v$\,sin\,$i$ (top) and rotation period (bottom). Red and black symbols stand for our targets listed in Table \ref{Table3} and the literature objects in Table \ref{Table4}. Binaries studied in this work are labeled. The green dashed and dotted curves in the top panel indicate the break-up velocity for an ultra-cool dwarf with 60 and 15 M$_{\rm Jup}$, respectively.}
              \label{incvsini}
    \end{center}
    \end{figure}

\subsection{Determination of the radius and inclination axis angle.}\label{radius}

A lower limit on the radius of the targets (R\,sin\,$i$) can be obtained by combining $v$\,sin\,$i$ and T$_{\rm rot}$ in equation \ref{eq1}. We derived the R\,sin\,$i$ of our targets with period detections (in the case of LP 329$-25$AB and LP 415$-$20AB, we used the observed period and the individual $v$\,sin\,$i$ of each component), these are tabulated in column 7 of Table \ref{Table3}. In addition, we adopted the theoretical radii expected for our targets (Sections \ref{single} and \ref{binary}) to derive the likely inclination angle range of their rotation axis. These values are listed in columns 8 and 9 of Table \ref{Table3}, where 0 and 90 deg stand for an ultra-cool dwarf seen at its pole and equator, respectively. We also searched in the literature for those dwarfs with measurements of both T$_{\rm rot}$ and $v$\,sin\,$i$, and computed their R\,sin\,$i$ values. For those objects believed to be older than 500 Myr, we derived the inclination of their rotation axis by assuming a radius of 0.1 $R_{\sun}$. The name of the dwarfs, their spectral types, T$_{\rm rot}$, $v$\,sin\,$i$, references to the literature works, and their estimated R\,sin\,$i$ are listed in Table\,\ref{Table4}. 

In Figure \ref{incvsini} we plotted the estimated inclination values for our targets and those of the literature as a function of their measured $v$\,sin\,$i$ (top) and rotation period (bottom). It is interesting to note that all the observed ultra-cool dwarfs with $v$\,sin\,$i\ge30$ km s$^{-1}$ tend to have rotation axis with inclinations equal to or greater than 40 deg (top panel). An estimation of the break-up speed of a mature ($R\approx0.1$ $R_{\sun}$) ultra-cool dwarf can be computed from the expression $v\,=\,\sqrt{GM/R}\approx\,42.6\sqrt{M({\rm M_{Jup}})}$ km s$^{-1}$; we plotted this velocity for ultra-cool dwarfs with 60 and 15 M$_{\rm Jup}$ for different inclinations in the top panel of Figure \ref{incvsini}. These curves show that fast rotators with inclinations $\le30$ deg and $v$\,sin\,$i\ge30$km s$^{-1}$ can exist. However, their absence in this region can be likely an observational bias due to the difficulty of measuring variability in objects with low inclinations. Similarly, in the bottom panel of Figure \ref{incvsini}, there are no detections for objects with rotation periods greater than 4 h and inclinations lower than 40 deg, which is in agreement with the recent findings by \citet{2017ApJ...842...78V} for L3.5-T1.5 dwarfs with $J$-band variability. On the contrary, objects with rotation periods between $\approx$1 and $\approx$3 h are detected in all the range of inclinations.

 \begin{figure}
   \begin{center}
    \includegraphics[width=0.49\textwidth]{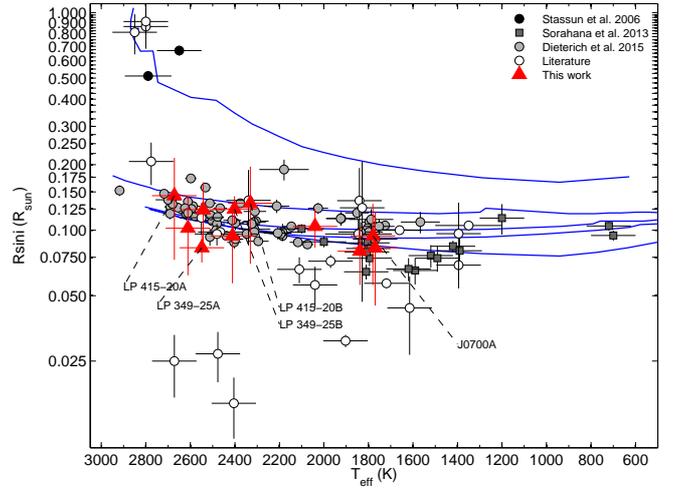}
     \caption{Lower limits on radius (R\,sin\,$i$) as a function of effective temperature (T$_{\rm eff}$). Red triangles stand for R\,sin\,$i$ estimated with the rotation periods presented in this work, while white circles are R\,sin\,$i$ derived with rotation periods taken from the literature (Table\,\ref{Table4}). Binaries studied in this work are labeled. Radii measurements (R) taken from the literature are plotted with gray circles \citep{2014AJ....147...94D}, gray squares \citep{2013ApJ...767...77S}, and black circles  \citep{2006Natur.440..311S}. The evolutionary models of \citet{2003A&A...402..701B} are also plotted (blue lines, from top to bottom: 1 Myr, 120 Myr, 500 Myr, 1 Gyr, and 10 Gyr). Uncertainties in the T$_{\rm eff}$ are $\pm100$ K.}
              \label{teff}
     \end{center}
    \end{figure}

We used the T$_{\rm eff}$-spectral type calibration given in \citet[][ compatible with those of \citealt{2013ApJS..208....9P} and \citealt{2013A&A...556A..15R}]{2009ApJ...702..154S} to convert spectral types to effective temperatures, and plotted the  values of R\,sin\,$i$ against T$_{\rm eff}$ in Figure\,\ref{teff}. Radii (R) measurements taken from \citet{2013ApJ...767...77S} and \citet{2014AJ....147...94D} are also plotted. As seen in this Figure, the derived R\,sin\,$i$ for our targets (red triangles) are in agreement with the evolutionary predictions at ages older than $\sim120$ Myr, and also with the radii measurements  for objects with similar spectral type and colors. This gives us confidence in the results obtained for the inclination angle distribution (based in the assumption of 0.1 $R_{\sun}$ for objects older than 0.5 Gyr). Remarkably, this way to obtain lower radii limits, or equivalently upper limits on the age, can be useful to set constraints on the evolutionary model predictions at very young ages, a range which is still poorly studied.

  \begin{figure}
   \begin{center}
  \includegraphics[width=0.49\textwidth]{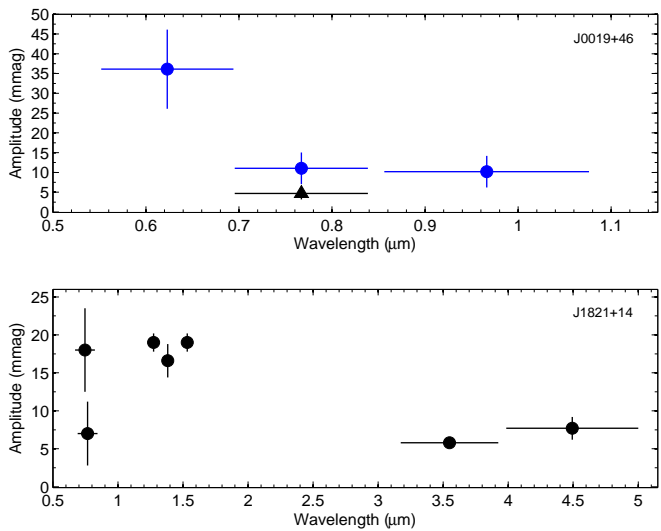}
     \caption{Observed photometric amplitude of variability as a function of the wavelength of observation for J0019$+$4614 (top) and J1821$+$1414 (bottom). Blue circles stand for simultaneous measurements. Vertical bars stand for the uncertainties of the amplitude. Horizontal bars stand for the filter passband. Data in the same filter have been slightly shifted for clarity.}
              \label{ampwav}
     \end{center}
    \end{figure}

\subsection{Amplitude of the photometric variability as a function of wavelength}

Two targets in our sample (J0019$+$4614 and J1821$+$1414) have photometric monitoring in different filters at optical and infrared wavelengths, allowing us to probe different atmospheric depths. 

In the case of J0019$+$4614, we collected simultaneous data in the Sloan $r$, $i$, and $z$ filters for nearly $\sim$2.5 continuous rotations ($\approx$5.5 h; Figure \ref{curvesIAC}). The behavior of the light curves in the three filters is very alike and does not show any significant phase lag between the peaks an valleys in the different filters. The measured amplitudes of variability are plotted as a function of the wavelength of observation in the top panel of Figure \ref{ampwav} (blue circles). Photometric amplitudes at $i$- and $z$- bands are almost identical ($\sim$10 mmag), and much smaller than the one measured in the $r$- band by a factor of $\sim$3. Different groups \citep[see ][]{1997ApJ...491..856B,2006ApJ...640.1063B,2001ApJ...556..872A} have investigated the effects of cloud opacity in the emergent flux of ultra-cool dwarfs, using particles of different composition and sizes. According to their models, the $r$-band flux of an object like J0019$+$4614 originates in layers of the atmosphere that are located deeper than those at which the silicates clouds are formed, while the $i$- and $z$-band fluxes come from regions located in the intermediate and upper parts of the cloud deck. As a result, the $r$-band flux will be more affected as it pass through the cloud cover and will show a larger photometric variability than fluxes at $i$- and $z$-bands. Thus, the difference between the variability amplitudes of the $r$- and the $i$- and $z$-bands gives a hint on the atmospheric structure of J0019$+$4614. We also included a previous measurement with the LT in the $i$-band (black triangle). This was measured from $\sim$4 continuous rotation cycles and is smaller than the other $i$-band amplitude by a factor of $\sim2.4$. Our measurements confirm the periodicity of this source in two epochs separated by 1.97 yr, and suggest that the different $i$-band amplitudes can be produced by a change in the size of the source of variability in the atmosphere of J0019$+$4614.

For J1821$+$1414, we plot in the bottom panel of Figure \ref{ampwav}: the $i$-band dispersion of its light curves presented  in this work, three $J$-band amplitudes of variability  reported by \citet[][]{2015ApJ...798L..13Y}, and two amplitude measurements at 3.6 and 4.5 $\mu$m \citep{2015ApJ...799..154M}. Again we find different amplitudes in the $i$-band at different epochs, which suggest a change in the source of variability; the largest amplitude in the $i$-band is similar to the ones observed in the $J$-band and much larger than the observed at 3.6 and 4.5 $\mu$m. However, simultaneous multi-wavelength measurements are needed since the variability source can be different from one epoch to another. These in combination with state-of-the-art radiative transfer codes have shown their utility to discern the nature (e.g., magnetic or dusty) of the observed photometric variability \citep{2013ApJ...767..173H,2015ApJ...813..104G}.


\section{Linear polarization and rotation}\label{sec4}

\begin{figure}
   \begin{center}
  \includegraphics[width=0.49\textwidth]{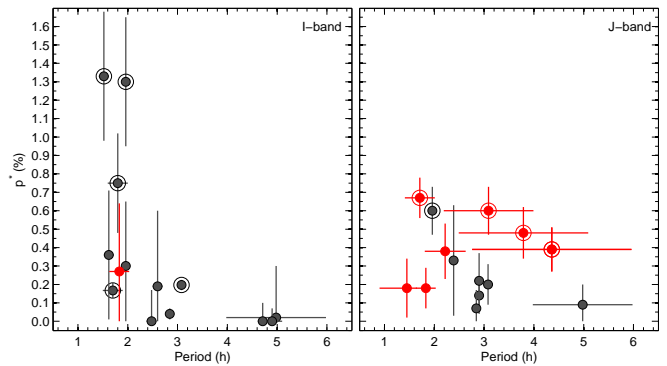}
     \caption{Degree of linear polarization in the $I$-band (left) and $J$-band (right) as a function of the rotation period. Red circles stand for our targets listed in Tables \ref{Table1} (polarization measurements) and Table\,\ref{Table3} (periods), while the gray ones denote data taken from the literature and listed in Table\,\ref{Table4}.}
              \label{pol}
     \end{center}
    \end{figure}

Ultra-cool dwarfs with dust in their atmospheres are predicted to exhibit non-zero linear polarization (see Section \ref{sec1}). Here we combine the published linear polarization measurements and rotation periods of our targets, and others from the literature, to investigate the likely relation between the occurrence of strong linear polarization and fast rotation at optical and near-infrared wavelengths. In Figure\,\ref{pol}, we plotted the optical and near-infrared degree of linear polarization ($p^{*}$) as a function of the rotation period for our targets and objects from the literature  listed in Tables\,\ref{Table1}, \ref{Table3} and \ref{Table4}. We note that near-infrared data (Figure\,\ref{pol}, right) were collected with the same instrumental setup, while optical data (Figure\,\ref{pol}, left) come from different instruments. 

To test if large values of linear polarization are associated to the fastest rotators, we searched for a correlation in the data of Figure \ref{pol}. Thus, we simply computed the Pearson-$r$ coefficient of linear correlation, obtaining $r$ values of -0.51 ($I$-band) and -0.21 ($J$-band) for the whole sample. When considering only significant measurements of linear polarization (i.e., $P/\sigma_{P}\ge3$; encircled symbols in Figure \ref{pol}), we find $r_{I}=-0.58$ and $r_{J}=-0.93$. The negative sign of the $r$ coefficients indicates that if there is a correlation, this will be between short rotation periods (fast rotations) and large degrees of linear polarization as previously pointed by \citet{2010ApJ...722L.142S} and \citet{2013A&A...556A.125M}. Near-infrared values of polarization with $P/\sigma_{P}\ge3$ show some kind of correlation, while the low value of $r$ derived from the optical measurements of linear polarization can be affected by using different $i$ filters and instruments. It should be noted that the dependency of linear polarization with rotation period (or oblatness) is not linear (e.g., see figure 1 in \citealt{2010ApJ...722L.142S} or figure 9 in  \citealt{2011MNRAS.417.2874M}), in particular, we may not be sampling properly the values of linear polarization for objects with rotation periods $\ge$5 h.

Despite the fact that the number of measurements may be small for rotation periods $\ge$5 h, the panels in Figure \ref{pol} also show that the very fast rotators (T$_{\rm rot}\lesssim2$ h) tend to exhibit a larger dispersion in the observed values of $I$-band linear polarization. Most of these measurements were collected in exposure times much shorter than the rotation period of the ultra-cool dwarf, thus, they represent the linear polarization state of a certain face of the ultra-cool dwarf. If the ultra-cool dwarf contains atmospheric heterogeneities, different values of linear polarization are expected with rotation as predicted by \citet{2011ApJ...741...59D}, and reported by \citet{2009A&A...502..929G} and \citet{2015A&A...580L..12M}. The surface heterogeneities of the dust clouds seem to play a role in the linear polarimetric variability detected in several ultra-cool dwarfs independently of their rotational velocity. From data presented in Figure \ref{pol}, it seems that fast velocities favor the detection of significant linear polarization at optical and near-infrared wavelengths as well as variability, but the net degree of linear polarization at a given epoch is mostly determined by the dust distribution on the surface of the ultra-cool dwarf.


\section{Conclusions}\label{sec5}

We have used the IAC80 and the LT to monitor a sample of 17 M7--L6 ultra-cool dwarfs in the $I$-band, and the WHT to monitor a L9.5 dwarf in the $J$-band. Targets studied in the $I$-band  were monitored continuously during 4 h in several ocassions (2-6). In each epoch, we derived the differential light curves for our targets obtaining photometric accuracies in the range  $\pm$1.5--8 mmag, with the exception of the very faint ($I\sim$20 mag) KPNO-Tau-4 and J0241$-$0326, for which we obtained accuracies of $\pm$30 mmag.

We compared the observed dispersion of the light curves of our targets with their associated photometric error to look for variability. We found that at a 2-sigma level (i.e., $\sigma_{\rm obs}/\sigma_{\rm err}\ge2$) 9 out of 18 targets are variable, 6 of them showing periodic light curves and 3 of them showing stochastic light curves. When we used periodogram algorithms to look for periodicities in our data, we found significant peaks in 7 of these objects. In addition, other 2 targets, which had not passed the 2-sigma criterion, showed significant peaks. We attribute these periodicities to rotation. The fraction of variable objects found in this work (50\%) is similar to the reported ones in the infrared by other groups. The measured rotation periods are in the range $\sim$1.5-4.4 h, which is the typical range of periods expected in field ultra-cool dwarfs based on their predicted sizes and measured $v$\,sin\,$i$.

We combined our detections of variability with others reported in the literature and found that  65\,$\pm$\,18 $\%$ of M7--L3.5 dwarfs with $v$\,sin\,$i\ge30$ km s$^{-1}$ are photometrically variable in the $I$-band. We also derived the most likely range for the inclination angle of the rotation axis of our targets and objects from the literature, and found that variable ultra-cool dwarfs with $v$\,sin\,$i\ge30$ km s$^{-1}$ are more likely to have inclination angles $\gtrsim40$ deg.

One of our targets (J0019$+$46) has simultaneous light curves covering $\sim$2.5 rotations in the Sloan $r$, $i$, and $z$ filters. $i$ and $z$ amplitudes are similar and smaller than the observed in $r$ by a factor of $\sim$3, which suggests some kind of atmospheric vertical structure. In addition, the observed $i$-band amplitudes in different epochs are different, suggesting a change in the size/location of the atmospheric spots of this dwarf.

Finally, we investigated whether fast rotators tend to exhibit larger values of linear polarization at optical and near-infrared wavelengths as predicted by the theory. We found that there is a correlation between short rotation periods and the detection of large degrees of linear polarization, when considering objects with significant measurements (i.e., $P/\sigma_{P}\ge3$).
\section*{Acknowledgements}

{\sl \small We are thankful to the anonymous referee for his/her valuable comments. We also thank the group of support astronomers at the LT for their valuable help in the preparation of the observations and the data adquisition. P. A. Miles-P\'aez thanks the telescope operators C. Mart\'inez-Lombilla and A. Bereciartua for their valuable help during some of the observing campaigns at the IAC80 telescope. The Liverpool Telescope is operated on the island of La Palma by Liverpool John Moores University in the Spanish Observatorio del Roque de los Muchachos of the Instituto de Astrofisica de Canarias with financial support from the UK Science and Technology Facilities Council. The IAC80 telescope is operated on the island of Tenerife by the Instituto de Astrof\'isica de Canarias in the Spanish Observatorio del Teide. This work is partly financed by the Spanish Ministry of Economics and Competitiveness through projects AYA2013-48391-C4-2-R, AYA2014-57495-C2-1-R, ESP2014-57495-C2-1-R, and ESP2016-80435-C2-2-R.}




\bibliographystyle{mnras}
\bibliography{biblio} 




\appendix

\section{Light curves of the reference stars for each epoch}\label{ap1}
\begin{figure*}
    \begin{center}
    \includegraphics[width=0.99\textwidth]{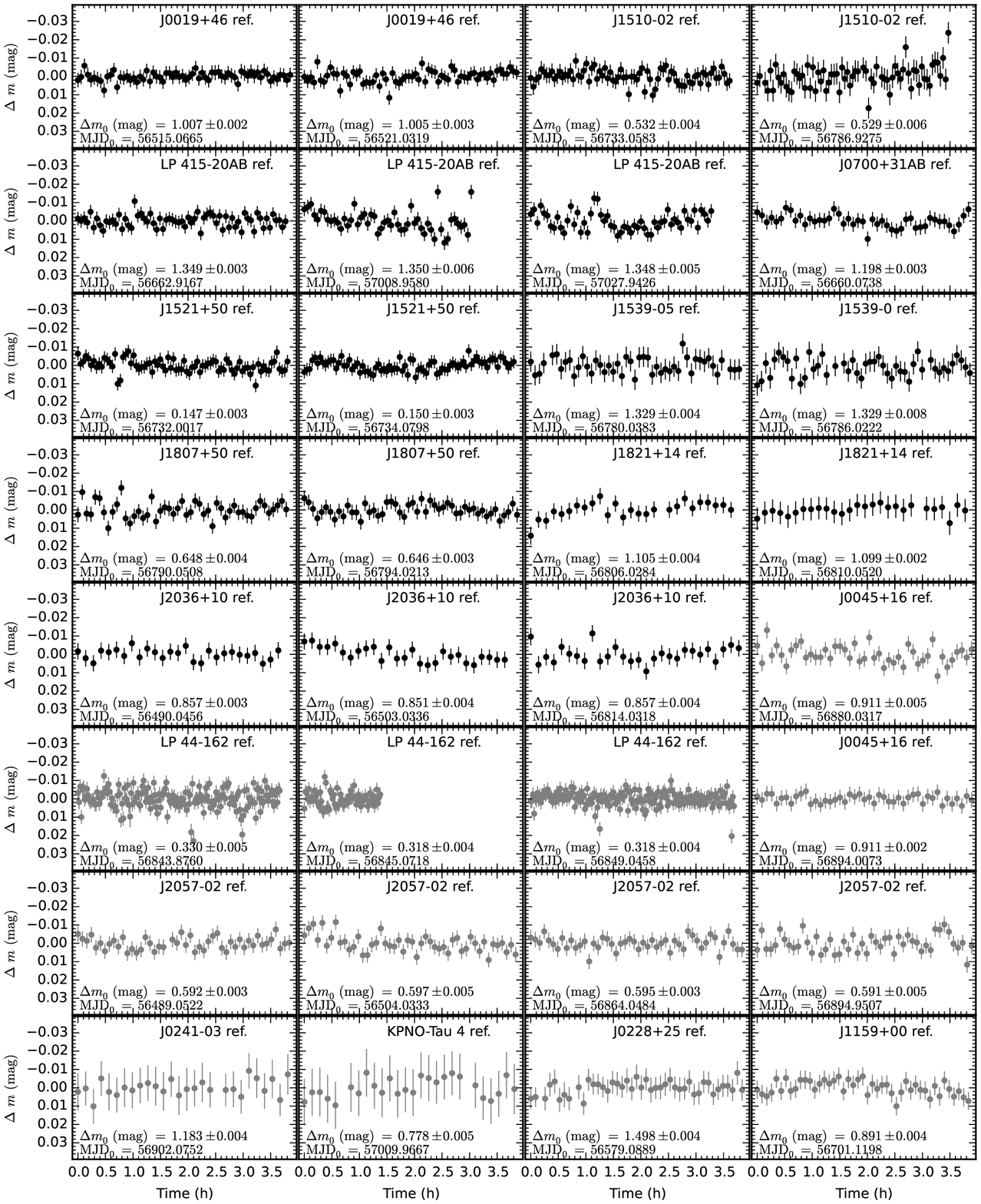}
     \caption{Individual LT light curves for one reference star with similar brightness to the target shown in each panel of Figure \ref{curvesLT}, vertical bars stand for photometric errors. The average value of the light curve ($\Delta$m$_{\rm o}$), its standard deviation, and the Modified Julian Date for time zero (MDJ$_{\rm o}$) are also given in all the panels. For comparison purposes, all the light curves have been shifted to the same zero-point, positive and negative values indicate darkening and brightening, respectively. Vertical and horizontal scales are the same as those displayed in Figure \ref{curvesLT}.}
              \label{curvesLTref}
   \end{center}
    \end{figure*}

\begin{figure*}
   \begin{center}
    \includegraphics[width=0.99\textwidth]{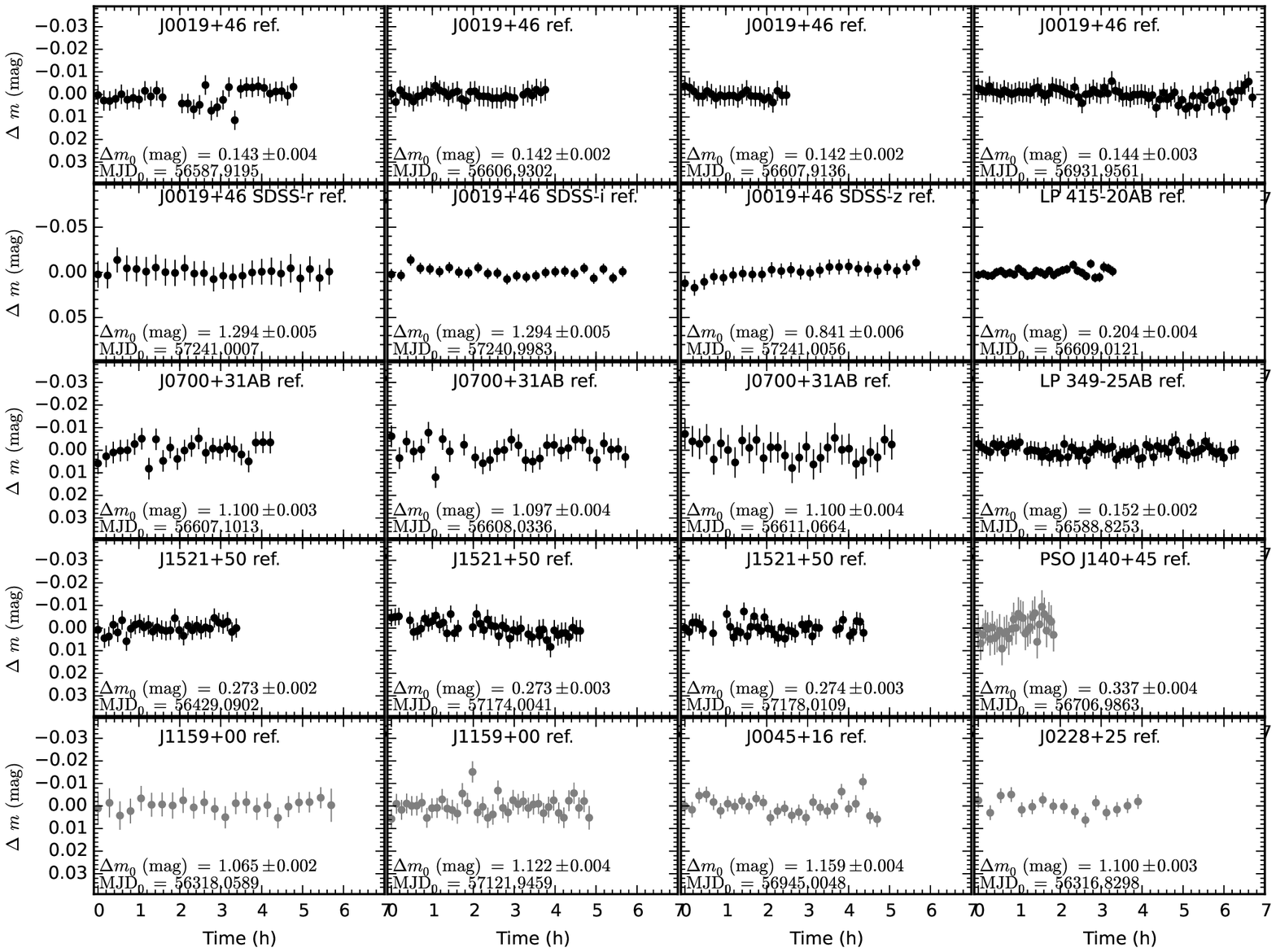}
     \caption{Individual IAC80 light curves for one reference star with similar brightness to the target shown in each panel of Figure \ref{curvesIAC}. For comparison purposes, vertical and horizontal scale are the same as those displayed in Figure \ref{curvesIAC}. Symbols as explained in Figure\,\ref{curvesLT}.}
              \label{curvesIACref}
     \end{center}
    \end{figure*}

Light curves for each single epoch at the LT and the IAC80 are plotted in Figures \ref{curvesLTref} and \ref{curvesIACref}, respectively.  If possible, we show the light curve of a comparison star with brightness similar to the target's and the same vertical and horizontal scale as in Figures \ref{curvesLT} and \ref{curvesIAC}. 


\bsp	
\label{lastpage}
\end{document}